 \documentclass[preprint2]{aastex}

\shorttitle{Chandra variability survey of M31 GCs}
\shortauthors{Barnard et al.}

\begin{document}


\title{12 years of X-ray variability in M31 globular clusters, including 8 black hole candidates, as seen by Chandra}


\author{R. Barnard and  M. Garcia}
\affil{Harvard-Smithsonian Center for Astrophysics, 60 Garden St, Cambridge MA 02138}
\and
\author{S. S. Murray}
\affil{Johns Hopkins University, Baltimore, Maryland}


\begin{abstract}
We examined 134 Chandra observations of the population of X-ray sources associated with globular clusters (GCs) in the central  region of M31.  These are expected to be X-ray binary systems (XBs), consisting of a neutron star or black hole accreting material from a close companion.  We created long-term lightcurves for these sources,  correcting for background, interstellar absorption and instrumental effects. We tested for variability by examining the goodness of fit for the best fit  constant intensity. We also created structure functions (SFs) for every object in our sample, the first time this technique has been applied to XBs. 
We found  significant variability in 28 out of  34 GCs and GC candidates; the other 6 sources had  0.3--10 keV luminosities fainter than $\sim$2$\times$10$^{36}$ erg s$^{-1}$, limiting our ability to detect similar variability. The SFs of XBs with 0.3--10 keV luminosities $\sim$2--50$\times10^{36}$ erg s$^{-1}$ generally showed considerably more variability than the published ensemble SF of AGN.  Our brightest XBs were mostly consistent with the  AGN SF; however, their 2--10 keV fluxes could be matched by $<$1 AGN per square degree.
 These encouraging results suggest that examining the long term lightcurves of other X-ray sources in the field may provide an important  distinction between X-ray binaries and background galaxies, as  the X-ray emission spectra from these two classes of X-ray sources are similar.
Additionally, we identify 3 new black hole candidates (BHCs) using additional XMM-Newton data, bringing the total number of M31 GC BHCs to 9, with 8 covered in this survey.    
\end{abstract}





\section{Introduction}

When studying the X-ray populations of external galaxies, globular cluster X-ray sources are of particular interest, because they are most probably low mass X-ray binaries (LMXBs), rather than e.g. background galaxies, supernova remnants, or high mass X-ray binaries (HMXBs). Globular clusters contain $\sim$10\% of the Galactic LMXB population (14 out of $\sim$150), but are thought to contain only 0.1\% of  the mass, hence  the LMXB density is $\sim$100 times higher in GCs than elsewhere \citep[see the comprehensive review by][ and references within]{vl06}; indeed, it has been speculated that all LMXBs were born in globular clusters \citep[see e.g.][ and references within]{dantona07}.

M31 is the nearest neighboring spiral galaxy, at $\sim$780 kpc \citep{sg98}, and is a favored target of many X-ray telescopes. Hence, its GC X-ray sources have been particularly well studied. X-ray surveys of M31  include those made with Einstein \citep{tf91}, ROSAT \citep{s97,s01}, XMM-Newton \citep{trudolyubov04,sg09, stiele11}, and Chandra \citep{kong02, distefano02, williams04}. 

\citet{distefano02} conducted a GC survey of M31 with Chandra, covering $\sim$2560 arcmin$^2$. They found 30 GC X-ray sources, 15 of which were new discoveries. They found that the M31 GC X-ray population was systematically more luminous than the Milky Way (MW) population; 10 M31 GC X-ray sources in their survey  exceeded 10$^{37}$ erg s$^{-1}$, vs. 1 of the  11 MW GC X-ray sources surveyed by ROSAT \citep{verbunt95}. 

Recently, \citet{peacock10b} conducted a survey of X-ray sources in the 2XMMi catalog \citep{watson09}  that included all publicly available XMM-Newton observations of M31, searching for associations with GCs; the observations cover $\sim$80\% of the GCs in M31. They found 41 GC associations in the 2XMMi catalogue , and 4 transient X-ray sources observed by either ROSAT or Chandra, for a total of 45 GC LMXBs. 

We have been monitoring the central region of M31 with Chandra for the last $\sim$12 years, averaging $\sim$1 observation per month, looking for X-ray transients; we exclude periods where M31 is too close to the Sun to be observed.  These observations have no fixed roll angle, and the combined field may be approximated by a circle with a $\sim$20$'$ radius.
The earlier Chandra surveys included long-term variability studies: \citet{kong02} found 204 X-ray sources in the central 17$'\times 17'$ region, detecting variability in $\sim$50\% of the sources over 8 ACIS-I observations between 1999, October and 2001, June; \citet{williams04} examined 2.5 years of HRC data, and found at least  25\% of 166 sources to be significantly variable.

In this work we present a variability survey of all X-ray sources in our Chandra observations that are associated with objects in the Revised Bologna Catalog of M31 globular clusters \citep[RBC V.4,][]{galleti04,galleti06,galleti07,galleti09}.  This  includes 30 out of 45 GC X-ray sources identified by \citet{peacock10b}; the other 7 RBC associations in our survey include 4 GC candidates, 2 background galaxies (AGN), and 1 star. We compare the variability of these X-ray sources with the ensemble variability for active galaxies found by \citet{vagnetti11}.

 It has long been known that AGN may vary by a factor of a 2--3 on time-scales of months to years, with the amplitude of variation inversely proportional to the luminosity \citep[see e.g.][ and references within]{marshall81,nandra97}.  However, it has been known for exceptional AGN to flare up by an order of magnitude \citep[e.g.][]{tan78}.

 Recently \citet{vagnetti11} studied the ensemble variability over time-scales of hours to years of AGN in the serendipitous source catalogs from XMM-Newton \citep{watson09} and Swift \citep{puccetti11}; their sample covered redshifts $\sim$0.2--4.5, and  0.5--4.5 keV luminosities $\sim$10$^{43}$--10$^{46}$ erg s$^{-1}$.  They included 412 AGN From the XMM-Newton catalog, and 27 AGN from the Swift catalog; all of these AGN were sampled at least twice. They used a structure function (SF) to estimate the mean intensity deviation for data separated by time $\tau$:
\begin{equation}
SF\left( \tau \right) \equiv \sqrt{\frac{\pi}{2}\left<|\log f_{\rm X} \left( t+\tau \right) - \log f_{\rm X} \left(t\right)| \right>^2 - \sigma_{\rm n}^2},
\end{equation}
where $\sigma_{\rm n}$ is the photon noise and $f_{\rm X}$ is the X-ray flux. They grouped  the SF into logarithmic bins with width 0.5; each bin in the range log($\tau$) = 1.0--3.0 contained more than 100 measurements.

 \citet{vagnetti11} found that SF increased from $\sim$0.1 to $\sim$0.2 as $\tau$ ranged over 0.1--1000 days in the rest frame of the AGN. SF $\sim$ 0.15 for $\tau$ = 30 days, suggesting that typical AGN would vary by $\sim$25--75\% between our observations. They also investigated the well-observed anti-correlation between intensity variability ($I_{\rm var}$)  and luminosity, expressed  in the form $I_{\rm var} \propto L_{\rm X}^{-k}$; $k$ $\sim$0.3 in the literature for time-scales of days to tens of days. They measured $k$ for AGN grouped logarithmically over log $L_{\rm X}$ = 43.5--45.5 for two values of $\tau$: 1 day and 100 days. They find $k$ = 0.42$\pm$0.03 for $\tau$ = 1 day, and $k$ = 0.21$\pm$0.07 for $\tau$ = 100 days. 

\citet{vagnetti11} measured the average values of  $\sigma_{\rm n}^2$ for the XMM-Newton and Swift samples to be 0.031 and 0.163 respectively. When they used these values to calculate the SFs for each sample, the two SFs were consistent; if SF($\tau$) $\propto$ $\tau^{b}$, then $b$ = 0.10$\pm$0.01 for the XMM-Newton SF, and $b$ = 0.07$\pm$0.04 for the Swift SF.

We have created structure functions for each of our targets from their 0.5--4.5 lightcurves.  The goal was to ascertain whether we may distinguish between XBs and AGN from their variability alone; this is important because XB and AGN often exhibit similar spectra.

Our analysis of the full data set  has allowed us to identify seven black hole candidates associated with globular clusters; four were previously published \citep{barnard11}, and three are new to this work. We identified one more GC black hole candidate in this region in earlier work \citep{barnard09}, and one away from the bulge \citep{barnard08}. They were all identified from their X-ray properties, as they exhibited characteristic low state emission spectra at luminosities that exceeded the limit for a neutron star accretor; other neutron star emission models were rejected \citep[see][ for the most complete justification of this method]{barnard11}.

 We describe the data analysis in the next section, followed by a presentation of our results, then a discussion of our findings and conclusions.

\section{Observations and data reduction}

 We have analyzed 89 ACIS observations and 45 HRC observations, in order to discern the variability of X-ray sources associated with GCs in this region.
The merged ACIS  image was registered to the Field 5 B band image from the Local Group Galaxy Survey observations of M31 \citep{massey06} with {\sc iraf} software, using  27 X-ray bright GCs.   After refining the  GC positions  in the Chandra and LGS B band  images with {\sc imcentroid}, we found the sky co-ordinates of the new positions with {\sc xy2sky}, then registered the X-ray positions to the LGS B band image using {\sc ccmap}. The mean r.m.s. uncertainty in position due to registration was 0.11$''$ in R.A. and 0.09$''$ in Dec.

We obtained 0.3--7.0 keV lightcurves and spectra from circular source and background regions for each source. The background region was the same size as the source region, and at a similar off-axis angle. The extraction radius varied between sources, because  larger off-axis angles resulted in larger point spread functions.

 For ACIS observations, we obtained corresponding response matrices and ancillary response files. Source spectra with $>$200 net source counts were freely fitted with absorbed power law models. For GCs with multiple freely fitted spectra, we obtained the best fit constant values for absorption ($N_{\rm H}$) and photon index ($\Gamma$).
 For observations with $<$200 net source photons, we fixed $N_{\rm H}$ and $\Gamma$ to the best fit values; if a particular GC had no freely fitted spectra, then we assumed $N_{\rm H}$ = 7$\times 10^{20}$ H atom cm$^{-2}$, and $\Gamma$ = 1.7. We obtained luminosities for faint sources by determining the unabsorbed 0.3--10 keV luminosity equivalent to 1 count s$^{-1}$;  multiplying the source intensity by this conversion factor gave the source luminosity, after correcting for the exposure, vignetting and background.

For HRC observations,  we included only PI channels 48--293,  thereby reducing the instrumental background. We used the WebPIMMS tool to find the unabsorbed luminosity equivalent to 1 count s$^{-1}$, assuming the same emission model as for the ACIS observations with $<$200 photons.  We created a 1 keV exposure map for each observation, and compared the exposure within the source region with that of an identical on-axis region, in order to estimate the necessary exposure correction. We multiplied the background subtracted, corrected source intensity by the conversion to get the 0.3--10 keV luminosity.

We created long term 0.3--10 keV  lightcurves for each source, using the luminosities obtained from each observation as described above. We only included observations with net source counts $\ge$ 0 after background subtraction. We fitted each long term lightcurve with a line of constant intensity, in order to ascertain the source variability. If the null hypothesis probability for the best fit to a lightcurve  was $<$0.003, then  the variability of that lightcurve had a significance $>$3$\sigma$, and we classified the target as variable; the null hypothesis probability determines the likelihood that observed deviations from the model are due to random fluctuations in the data.

 Luminosity uncertainties for freely fitted spectra were estimated with XSPEC by calculating a range of fluxes obtained by varying the emission parameters; the uncertainties for the faint spectra are derived directly from intensity uncertainties using the relevant  fixed model.

We derived SFs from the  0.5--4.5 keV fluxes of each observation of every target by assuming a power law spectrum with the same photon index as for the HRC and faint ACIS observations; an M31 X-ray source with a 0.3--10 keV unabsorbed luminosity of 1.0$\times 10^{37}$ erg s$^{-1}$ and $\Gamma$ = 1.7 has a 0.5--4.5 keV flux of 0.8$\times 10^{-13}$ erg cm$^{-2}$ s$^{-1}$.  

 \citet{vagnetti11} calculated the noise component from
\begin{equation}
\sigma_n^2 = 2\left<\left(\delta \log f_{\rm X}\right)^2\right> \simeq 2\left(\log e\right)^2 \left< \left( \frac{\delta f_{\rm X}}{f_{\rm X}}\right)^2\right>,
\end{equation}
assuming that $\delta f_{\rm X}/f_{\rm X}$ = $\left(1/N_{\rm phot}\right)^{0.5}$, and $N_{\rm phot}$ is the number of photons. Our lightcurves are background subtracted, and ARF-corrected; furthermore, uncertainties in the luminosities of bright sources include uncertainties in the spectral parameters.  As a result, our uncertainties are not simply due to photon counting noise. Hence in our case,
\begin{equation}
\sigma_n^2 \simeq \left(\log e\right)^2 \left< \left[ \frac{\delta f_{\rm X}\left(t+\tau\right)}{f_{\rm X}\left(t+\tau\right)}\right]^2 + \left[ \frac{\delta f_{\rm X}\left(t\right)}{f_{\rm X}\left(t\right)} \right]^2\right>.
\end{equation}

\section{Results}
\label{res}

The merged observations cover an approximately circular region with 20$'$ radius. We detected 430 X-ray sources in this region, and found 428 globular clusters from the RBC. Therefore we expect chance coincidences of X-ray sources within 1$''$ of the GC centers for 0.12 out of 428 GCs.

 Shifting the declination of each source by $\pm$5$''$ resulted in no coincidences within 1$''$ of a RBC object; shifting the RA by $-$5$''$ also resulted in no coincidences; however, shifting the RA by +5$''$ resulted in two matches within 1$''$ of a RBC object. Expanding the search radius from 1$''$ to 2$''$  provided one additional match each for offsets $\pm$5$''$ in declination. Hence we expect 0--2 false matches. 

 We found 37 X-ray sources to be coincident with M31 GCs, as classified by the RBC. X-ray positional uncertainties include uncertainties in the centroid on the X-ray image, uncertainties in the registration (0.11$''$ for RA, and 0.09$''$ for Dec), and 0.25$''$ uncertainty in the LGS position.

In Table~\ref{opticalprops} we present the positions and identifications of each X-ray source that we associate with a member of the RBC V4. We provide the latest classifications of the optical counterparts, gleaned from \citet{caldwell09}; our sample includes 1 star, and  2 galaxies, as well as 30 confirmed old GCs and 4 GC candidates. We also show the distance between the X-ray source and its optical counterpart, and the uncertainties in RA and Dec; X-ray sources with no good centroid solution are indicated by ``$\dots$''.  
Eight of our targets are far off-axis ($\sim$11--20$'$) and also X-ray faint. We estimated the positional uncertainties for these targets by binning the merged image by a factor 9 in $x$ and $y$, then running {\sc imcentroid} on this binned image.

%

Table~\ref{xrayprops} details the X-ray properties of each target. We first give the number of Chandra observations with $\ge$0 net source counts. We then show the line of sight absorption and photon index used for the model used to convert from intensity to luminosity for all HRC observations, and for ACIS observations with $<$200 net source photons. We finally give the best fit 0.3--10 keV unabsorbed luminosity and $\chi^2$/dof for the best fit line of constant intensity. The lightcurves of the variable sources are presented in Fig.~\ref{gclcs}; the complete version of this figure is available in the electronic version.

\subsection{Variability analysis}

We found significant X-ray variability in 28 of the 34 GCs and GC candidates; the remaining 6 sources had 0.3--10 keV luminosities $\la$2$\times 10^{36}$ erg s$^{-1}$. The two AGN in our survey, B042D and B044D were rather faint ($<2\times 10^{36}$ erg s$^{-1}$), but both exhibited variability at at $>3\sigma$ significance.

In Fig.~\ref{sfh} we present a histogram showing the number of pairs of observation with a given separation $\tau$ for XB144, the target with the most observations. The histogram is binned logarithmically, with a width of 0.2. Each bin averages tens or hundreds of observation pairs.

We note that SF($\tau$)  becomes imaginary if the observed variability is smaller than the noise. Such behavior manifests as SF($\tau$) = 0. Both AGN have mostly imaginary SFs; however, both show significant variability over times-scales of a few thousand days; this variability is consistent with the ensemble AGN SF created by \citet{vagnetti11}. 

The SFs of the brightest X-ray binaries (0.3--10 keV luminosity $\ga$5$\times 10^{37}$ erg s$^{-1}$) are also generally consistent with the ensemble AGN SF, or even less variable, despite the relatively high signal to noise. This result is in keeping with the observed behavior of Galactic X-ray binaries; \citet{muno02} showed that the brightest X-ray binaries (i.e. Z-sources) vary by a factor of a few, while the fainter binaries (atoll sources) can vary by 1 or 2 orders of magnitude. However, these bright X-ray sources have 2--10 keV fluxes greater than nearly all AGN; the 2--10 keV log N-log S plot for AGN calculated by \citet{moretti03} leads us to expect $<$1 AGN per square degree with comparable fluxes.

Our most encouraging result is that most of the fainter X-ray sources have SFs showing substantially more variability than the ensemble AGN SF over most time-scales. This is an important distinction to make, since X-ray binary and AGN emission spectra are often very similar. 

Unlike the other bright X-ray sources, the SF  for XB158 shows considerably more variability than the ensemble AGN  SF over all time-scales. XB158 is a high inclination X-ray binary that exhibits periodic intensity dips on a $\sim$10,000 s period\citep{trudolyubov02} and also disc precession \citep{barnard06}; it is further discussed in Section~\ref{b158}.

\subsection{Eight black hole candidates}
We have previously identified six black hole candidates (BHCs) in the observed region, based on their exhibition of characteristic low state emission spectra at luminosities that are too high for neutron star systems \citep{barnard08,barnard09,barnard11,barnard11b}. Five of these systems are associated with GCs, and four of those are persistently bright, which is unexpected given that much of the theoretical work to date models these systems as transients \citep[see ][ for an overview of the literature on the topic of black holes in globular clusters]{barnard11}.

Low state emission spectra are characterized by a dominant power law component with $\Gamma$ $<$2, contributing $\ga$90\% of the flux \citep{mr06}, and are seen in all low mass X-ray binaries, whether they contain black holes or neutron stars \citep{vdk94}. However, it has become apparent that the low state is confined to luminosities $\la$10\% Eddington \citep{glad07,tang11}. Hence X-ray binaries that exhibit low state spectra at luminosities  significantly higher than 3$\times 10^{37}$ erg s$^{-1}$ are likely to contain black hole accretors.

However, a  good fit with $\Gamma$ $<$2 is not sufficient by itself to identify a BHC; we must first reject competing emission models. In particular, we must reject a disk blackbody-dominated spectrum (representing the thermally dominated black hole state), a blackbody + power law model that represents high luminosity emission in  neutron star XBs, and a disk blackbody + steep power law ($\Gamma$ $>$2.4) model to represent the BH steep power law state. 

The  Terzan-5 X-ray transient IGR\thinspace J17480$-$2446  exhibited an unusually impressive range of properties: Z-source and atoll behavior, as well as thermonuclear X-ray bursts and 11 Hz pulsations;  \citet{chakraborty11} conducted  detailed spectral analysis of the outburst, using data from $\sim$40 $\sim$daily RXTE observations. They found that the spectra were all well described by a simple emission model (blackbody + power law + Gaussian emission line). The blackbody component contributed $\sim$30--50\% of the 3--15 keV flux, with k$T$ = 1.4--2.1 keV for the most part and 3.7 keV for the first observation; $\Gamma$ $\sim$2.1--3.0; the unabsorbed 3--15 keV flux was 2.58--17.94$\times 10^{-9}$ erg cm$^{-2}$ s$^{-1}$ \citep{chakraborty11}.

 We estimated the unabsorbed luminosity and blackbody contribution to each spectra in the 0.3--10 keV band for IGR\thinspace J17480$-$2446, from the best fit parameters, and assuming a distance of 5.5 kpc. We assumed that the power law component extended down to 0.3 keV; alternative models where the seed photon energy for the Comptonized component matched the inner disk temperature were rejected by the XMM-Newton spectrum of our  M31 black hole candidate RX J0042.3+4115 in the steep power law state \citep{barnard11b}.  We found that the blackbody contributed 7--13\% of the unabsorbed 0.3--10 keV luminosity of 1.8--33$\times$10$^{37}$ erg s$^{-1}$. IGR\thinspace J17480$-$2446 appears not to have been in the low state at any time during the outburst.

We note that \citet{lin07} devised an alternative model to describe two other neutron star X-ray transients, Aql X-1, and 4U\thinspace 1608$-$52; traditional emission models (thermal + inverse-Comptonized emission) all yielded good fits for spectra with $>$10$^{6}$ net source counts, but the authors raised objections to various aspects of the spectral evolution of the transient outbursts. They preferred a disk blackbody + blackbody emission model at high luminosities, with a Comptonized component necessary at lower luminosities. In this scenario, the neutron star transient evolution closely follows black hole transient evolution, with an extra blackbody component from either the neutron star itself or the boundary layer between the disc and neutron star surface. 

However, their model contradicts a large body of work based on observations of Galactic neutron star XBs. Plenty of evidence exists for an extended X-ray emission region in the high inclination, ``dipping'' X-ray binaries \citep[radius $\sim$20,000--700,000 obtained from ingress of photo-electric absorption dips; see][and references within]{cbc04}. Furthermore, broadened emission lines in Chandra observations of Cygnus X-2 suggest a hot, dense corona of up to $\sim$100,000 km \citep{schulz09}. Nevertheless, we applied the double-thermal emission model to our BHC spectra for completeness; the Comptonized component should not be required at such high X-ray luminosities according to \citet{lin07}.

 We examined the lightcurves and spectral histories of all our targets, searching for further BHCs.
We found a total of 8 GC LMXBs that apparently exhibited high luminosity low states; we refer to the X-ray source associated with GC labeled in the RBC v4 as B$nnn$ as XB$nnn$. In addition to the known BHCs  XB082, XB144, XB153, XB163, and  XB185 \citep{barnard09,barnard11}, we have identified XB086, XB135 and XB148 as new BHCs. We present the lightcurves and spectral histories ($\Gamma$ vs. time) for the 8 BHCs in Fig~\ref{bhlcs}; the complete figure is available in the electronic edition. The best fit blackbody + power law emission models for XB082, XB144, XB153, XB163, and XB185 are presented in Barnard et al. (2009, 2011); they clearly demonstrate the low state nature of the emission. 

In the following spectral analysis, uncertainties in best fit parameters and luminosities for XB086, XB135, and XB148 are quoted at the 90\% confidence level, corresponding to $\sim$1.6$\sigma$.

\subsubsection{XB086}
XB086 exhibited $\Gamma$ $\sim$1.4 at 0.3--10 keV luminosities $\sim$6--10$\times 10^{37}$ erg s$^{-1}$. Furthermore, it exhibited spectral variability $\sim$1600 days into the observations; $\Gamma$ changed from 2.0 to 1.5 accompanied by a luminosity decrease from 1.1$\times 10^{38}$ to 7.5$\times 10^{37}$ erg s$^{-1}$. This variation may indicate a state change  from e.g. the steep power law state to the low state. We note that XB086 apparently exhibited two different spectral states at very similar luminosities; hysteresis in the state transitions of black hole binaries where the transition from high state to low state is at a lower luminosity than the transition from low state to high state is well known \citep[see e.g.][]{maccarone03}.

None of the ACIS spectra were of sufficient quality to discriminate between spectral states, so we obtained a spectrum for XB086 from the 2002 January 6 XMM-Newton observation (ObsID 0112570101). This spectrum rejected a disk blackbody model (k$T_{\rm in}$ = 1.7 keV $\chi^2$/dof $=$284/155), but a power law model provided a good fit ($\Gamma$ = 1.48$\pm$0.05, $\chi^2$/dof = 155/154).

Adding a blackbody to the power law resulted in a slightly improved fit (k$T$ = 0.9$\pm$0.3 keV, $\Gamma$ = 1.47$\pm$0.18, $\chi^2$/dof $=$141/152); the blackbody component contributed $\sim$12\% of the 0.3--10 keV luminosity  ($\sim$6$\times 10^{37}$ erg s$^{-1}$). The Terzan 5 NS transient exhibited k$T$ $\sim$2.0 and $\Gamma$ $\sim$2.3 at similar luminosities \citep{chakraborty11}; these parameters differ by $>$5$\sigma$. Fig.~\ref{b086spec} shows the best fit two component model to the unfolded  spectrum, multiplied by channel energy; it is representative of the BH low state, and inconsistent with a high luminosity NS XB. This spectrum containted $\sim$9000 net source counts. 

Fitting a blackbody + disk blackbody model to B086 resulted in a 1.52$^{+0.18}_{-0.13}$ keV blackbody and an inner disk temperature of 0.66$\pm$0.08; the blackbody component contributed 64\% of the unabsorbed flux. The inner disk temperature obtained for XB086 is $>$5$\sigma$ below the minimum temperatures observed by \citet{lin07}; furthermore, \citet{lin07} never observed blackbody contributions $>$50\% in any of their transient spectra. Therefore the B086 spectrum is quite unlike any of the spectra observed by \citet{lin07}.

 We conclude that XB086 was in a low state; $N_{\rm H}$ = 1.06$\pm$0.15$\times 10^{21}$  atom cm$^{-2}$, and the unabsorbed 0.3--10 keV luminosity was 6.4$\pm$0.2$\times 10^{37}$ erg s$^{-1}$ (90\% confidence limits). This result supports our BHC classification from the ACIS observations.  

\subsubsection{XB135}
XB135 exhibited $\Gamma$ $\sim$1.6 at incredibly high luminosities: $\sim$4--6$\times 10^{38}$ erg s$^{-1}$ during the Chandra observaations. The highest quality spectrum (ObsID 14198) had a $\sim$40 ks exposure time, and a count rate $\sim$0.2 count s$^{-1}$. This high intensity would result in significant pile-up if XB135 were on-axis; however, it is near the edge of the field of view, and the photons are distributed over a large number of pixels. Indeed, no pixel accumulated more than 19 photons over 40 ks, hence pile-up was not an issue. 

The ObsID 14198 0.3-7.0 keV XB135 emission spectrum was well described by an absorbed power law ($\Gamma$ = 1.72$\pm$0.08, $\chi^2$/dof = 205/221), a disk blackbody (k$T_{\rm in}$ = 1.63$\pm$0.08 keV, $\chi^2$/dof = 197/221, or a blackbody + power law model (k$T$ = 0.7$\pm$0.2 keV, $\Gamma$ = 1.5$\pm$0.4, $\chi^2$/dof = 187/219).

We have previously published analysis of an XMM-Newton observation of XB135 \citep{barnard08}. It was well described by a power law emission model ($\Gamma$ = 1.56$\pm$0.03, $\chi^2$/dof = 467/435, but the fit was significantly improved by adding a thermal component (k$T$ = 0.8$\pm$0.2, $\Gamma$ = 1.54$\pm$0.14, $\chi^2$/dof = 413/433); the thermal component contributed 11$\pm$5\% of the flux. A disk blackbody-dominated model was rejected. We present the best fit blackbody + power law emission model to the pn spectrum in Fig.~\ref{b135spec}; this spectrum contains  $\sim$15000 net source counts.

The Terzan 5 transient exhibited kT = 1.64$^{+0.05}_{-0.07}$ keV and $\Gamma$ = 2.76$^{+0.06}_{-0.08}$ at its peak flux \citep{chakraborty11}; these are extremely different from the observed parameters for B135. Fitting a blackbody + disk blackbody model resulted in a 1.8$^{+0.8}_{-0.3}$ keV blackbody contributing 49\% of the 0.3--10 keV unabsorbed luminosity (3.7$\times 10^{38}$ erg s$^{-1}$); the inner disk temperature was 1.0$\pm$0.2 keV. Such a low disk temperature is consistent only with the lowest luminosities of the NS transients measured by \citet{lin07}, and  the blackbody contribution is too large. We conclude that this B135 spectrum is unlike either model for NS high states, and most liklely represents a low state BH XB.

The inferred mass for XB135 is very large, at least 33 $M_{\odot}$ from the XMM-Newton observation, and possibly $>$50 if XB135 was also in the low state during the ACIS observations. Such a massive black hole could have been formed from a metal poor high mass star, so that little mass was lost in stellar winds during the aging process \citep{belczynski10,mapelli10}, or could have been formed by stellar mergers in  the cluster core \citep{miller02}.

\subsubsection{XB148}

XB148 exhibited a hard emission spectrum ($\Gamma$ $\sim$1.5) up to a 0.3--10 keV luminosity of 8$\times$10$^{37}$ erg s$^{-1}$. Unfortunately, none of the ACIS spectra suggesting high luminosity low states were of sufficient  quality to discriminate between emission models. 

A 2008 January 27  XMM-Newton observation of XB148  (ObsID 0505720501) with $\sim$2000 net source counts yielded a 0.3--10 keV spectrum that was well described by a power law ($\Gamma$=1.80$\pm$0.15, $\chi^2$/dof = 96/115) or a disk blackbody (k$T_{\rm in}$ =  1.14$\pm$0.15, $\chi^2$/dof = 125/116); these results are consistent with a black hole low state or high state.
 Fitting a blackbody + power law model resulted in k$T$ = 0.4 keV, unexpectedly low for a neutron star system; we present this best fit to the unfolded spectrum times channel energy in Fig.~\ref{b148spec}.

The Terzan 5 transient exhibited k$T$ = 1.4--3.7 \citep{chakroborty11}, for 0.3--10 keV luminosities 2.5--33.7$\times 10^{37}$ erg s$^{-1}$, $>$5$\sigma$ higher than for XB148; similarly, $\Gamma$ for XB148 is $>$4$\sigma$ lower than the range ovbserved by \citet{chakraborty11}. Fitting a blackbody + disk blackbody model yielded k$T$ = 3.5 keV, with unconstrained uncertainties, contributing 42\% of the unrealistic 0.3--10 keV luminosity (1.3$\times 10^{39}$ erg s$^{-1}$); the inner disk temperature is 0.80$\pm$0.10, 3$\sigma$ below the lowest temperature measured by \citet{lin07}.

These results are supportive of the classification of XB148 as a black hole candidate.
 Assuming a power law model yields $N_{\rm H}$ =  1.5$\pm$0.4$\times 10^{21}$ atom cm$^{-2}$,  giving an unabsorbed 0.3--10 keV luminosity of 4.7$\pm$0.5$\times 10^{37}$ erg s$^{-1}$; hence, we cannot rule out a particularly massive ($\sim$2.8 M$_{\odot}$) neutron star accretor in its low state.

\subsubsection{Properties of the host clusters}

We now consider the properties of the GCs hosting our BHCs. In particular, we examine the masses, ages, and metallicities of these clusters; these values were obtained from \citet{caldwell11}, who also ranked the 379 M31 GCs in their study by mass and by metallicity. We list these properties in Table~\ref{gcprops}.

In \citet{barnard11} we concluded that particularly massive or metal rich GCs could contain bright X-ray sources (and hence BHCs) as well as those GCs that are both massive and metal rich. The GCs containing our three new BHCs are more massive than $>$80\% of the GCs surveyed by \citet{caldwell11}; however, two of them (B086 and B135) are in the lowest quartile for metallicity, with [Fe/H] = $-$1.82. We note that the low metallicity for B135 is in keeping with the high inferred BH mass for XB135; \citet{belczynski10} have shown that while BH masses are limited to $\sim$15 M$_{\odot}$ for solar metalicities, they could theoretically  reach $\sim$86 $M_{\odot}$ for metallicities $\sim$1\% Solar.  

\subsection{The dipping, precessing X-ray binary XB158, and related systems}
\label{b158}
XB158 (a.k.a. Bo 158) is a high inclination XB that experiences periodic intensity dips on the orbital period \citep{trudolyubov02}; the original study used lightcurves folded on the 10017  s (2.8 hr) period, and \citet{trudolyubov02} concluded that the dip depth varied with luminosity. However, detailed analysis of the original XMM-Newton observations, along with proprietary observations, revealed that the dips appeared in some observations, but not in others \citep{barnard06}. This behaviour occurs in  systems with low mass ratios (short periods) because the outer edge of the accretion disc reaches the 3:1 resonance, causing additional tidal torques from the secondary that lead to elongation and  precession of the disc \citep[see e.g.][]{osaki89,wk91,ogilvie01}. 

XB158 is located at a high off-axis angle from the focus of the observations, and was only observed in 19 ACIS observations; it was observed in all of the HRC observations thanks to the larger field of view. The 0.3--10 keV luminosity was extremely variable, ranging over $\sim$4--40$\times 10^{37}$ erg s$^{-1}$, with the best fit line of constant intensity yielding $\chi^2$/dof = 3030/63 for ACIS and HRC observations.

We note that the high quality XMM-Newton spectra of XB158 require a two component emission model consisting of  e.g. a black body + power law \citep{barnard06}, however, the Chandra spectra were well described by a power law with best fit constant  $\Gamma$ = 0.59$\pm$0.17.

 We propose that the observed fluctuation between high and low luminosities may be caused by varying accretion rates over the disc precession cycle. Using proprietary three dimensional smoothed particle hydrodynamics code, we estimated the disc precession period to be $\sim$29 orbital periods, or $\sim$81 hours \citep{barnard06}. The cadence of these observations is not well suited for testing such a period, however the similar minima and maxima are suggestive of some sort of super-orbital cycle. 

We note that the HRC luminosities for XB158 may be misleading, since we are approximating a two-component spectrum with a single power law, and the HRC responses to each component may be different from  the ACIS responses. However, it is encouraging to see that ACIS and HRC observations made at similar times produced consistent luminosities.  Even if the  normalization of the HRC lightcurve is incorrect, the variability reflects the intensity variation of the source, suggesting that the luminosity of XB158 really is rapidly, and systematically variable.

We may expect GC XBs to have systematically shorter periods than XBs outside GCs because the probability of collisions  with other stars in the GC is high, and these interactions shorten the orbital periods. Hence, even binaries with mass ratios $>$0.3 may have sufficiently short periods to experience the  tidal resonances that produce precessing discs. 

The 0.3--10 keV lightcurve of XB146 looks similar to that of XB158, in that it 
also fluctuates between fairly consistent minima and maxima; however, it is observed in 131 out of 134 observations, and appears to be more structured.  The best fit line of constant intensity yields $\chi^2$/dof = 3614/130, as the 0.3--10 keV luminosity fluctuates between $\sim10^{37}$ and $\sim$5$ \times 10^{37}$ erg s$^{-1}$, with the occasional outlier. We therefore suggest that XB146, and other X-ray sources exhibiting similar behavior are XBs with precessing accretion discs. Unfortunately, this is not easily verified unless they are viewed close to  edge on.

\subsection{X-ray transients}
Four X-ray sources in our sample are considered transient, in that they vary in 0.3--10 keV luminosity by a factor $>$100 between observations where they are detected, and are fainter than the detection limit in other observations. Two types of XB behaviour leads to transient X-ray sources. LMXBs may exhibit transient behaviour due to instabilities in the disc; the discs accumulate matter during a long quiescent phase, and burn a large portion of the accreted material during outbursts. Long period HMXBs with eccentric orbits may also be X-ray transients, if accretion occurs only in a limited phase range near periastron.

Three of the transients are associated with a GC or candidate and are expected to be LMXBs due to the ages of the GC populations; this includes XB163, which we have identified as a black hole candidate, as previously discussed.

 XB128 varied by a factor $>$500 over 47 observations, peaking at a 0.3--10 keV luminosity $\sim$5$\times 10^{37}$ erg s$^{-1}$. The best spectrum for XB128 was from Chandra observation 4682 (2004 May 23), yielding 157 net source counts; it was well described by a power law with $\Gamma$=1.5$^{+0.9}_{-0.7}$, a disk blackbody with k$T$ = 1.6$^{+0.9}_{-0.4}$ keV, or a two component model consisting of a $\sim$1 keV blackbody and a power law with $\Gamma$ $\sim$2; this spectrum is consistent with a neutron star or black hole primary.

 XPB-in7 varied by a factor $>$100 over 103 observations. The peak luminosity of XPB-in7 was in Chandra observation 8183 (2007 January 14), with 227 net counts; however, we were unable to obtain an acceptable fit from any of our spectral models. Since the peak luminosity was only 1.7$\times 10^{37}$ erg s$^{-1}$, we would be unable to determine the nature of the accretor in any case.  We note that PB-in7 is only a GC candidate, and further insight into the nature of the host may help interpret the nature of XPB-in7.


XSK059A is coincident with  likely compact GC, in a region rich in HII (Caldwell, private communication). As such, it may be a HMXB associated with the HII instead of a LMXB associated with the GC. XSK059A  varied by a factor $\sim$300 between Chandra observations, exhibiting long quiescent intervals punctuated with many outbursts. The lightcurve is consistent with 8 outbursts $\ga$10$^{37}$ erg s$^{-1}$ separated by 1--6 cycles of $\sim$120 days, with
the  outbursts lasting up to $\sim$90 days; we indicate these bursts in Fig. 1 with downward arrows. However, these outbursts are not present in every cycle. Such behaviour may be explained by a HMXB with variable mass loss rate from the secondary. 

\section{Discussion and conclusions}

We have analyzed the long term lightcurves of globular cluster X-ray sources using 134 Chandra observations taken over $\sim$12 years; these X-ray sources are likely LMXBs. Our sample was drawn from the Revised Bologna Catalogue V. 4, and  included 30 confirmed old globular clusters, 4 candidate globular clusters, 2 background galaxies, and 1 star. Fitting each lightcurve with a line of constant luminosity revealed that 30 out 37 targets were variable; the other 7 targets had 0.3--10 keV luminosities $\la$2$\times 10^{36}$ erg s$^{-1}$. We infer from these results that all XBs in our field with 0.3--10 keV luminosities $\ga2\times10^{36}$ erg s$^{-1}$ should be variable also.

We created structure functions from the 0.5--4.5 keV fluxes of our variable sources, for comparison with the ensemble AGN SF created by \citet{vagnetti11}. The lower luminosity XBs exhibited SFs that generally showed substantially more variability than the AGN SF over most time-scales.  The higher luminosity XBs had SFs generally indicating similar, or lesser, variability than the ensemble AGN SF; however, the 2--10 keV fluxes for these systems were higher than almost all AGN.  This discrimination is important, since the emission spectra of XBs and AGN are often similar.
 
We have identified a total of 9 black hole candidates in M31 GCs, of which 8 are covered by this survey; 7 BHCs were identified in these data (4 reported in Barnard et al., 2011, and 3 new to this work), using our well established method of seeking low state emission spectra at luminosities that conspicuously exceed the limit for neutron star binaries. We have identified apparent super-orbital variability in XB158, which is a high inclination system where the disk is known to precess. The apparently regulated behaviour of XB146 may also indicate a short period system with a precessing disc.

 We will be creating SFs for  the remaining $\sim$400 X-ray sources in our field, and expect to identify $\ga$100 new X-ray binaries.



\section*{Acknowledgments}
We thank the referee for their thoughtful comments that resulted in a much improved paper. We thank Z. Li for merging the Chandra data.  This research has made use of data obtained from the Chandra data archive, and software provided by the Chandra X-ray Center (CXC). We also include analysis of data from XMM-Newton, an ESA science mission with instruments and and contributions directly funded by ESA member states and the US (NASA).
R.B. is funded by Chandra grants GO0-11106X and GO1-12109X, along with  HST grants GO-11833 and GO-12014.  M.R.G. and S.S.M are  partially supported by NASA grant NAS-03060.



{\it Facilities:} \facility{CXO (ACIS)} \facility{CXO (HRC)}.



\bibliographystyle{aa}

\clearpage





\begin{figure*}
\epsscale{2.2}
\plotone{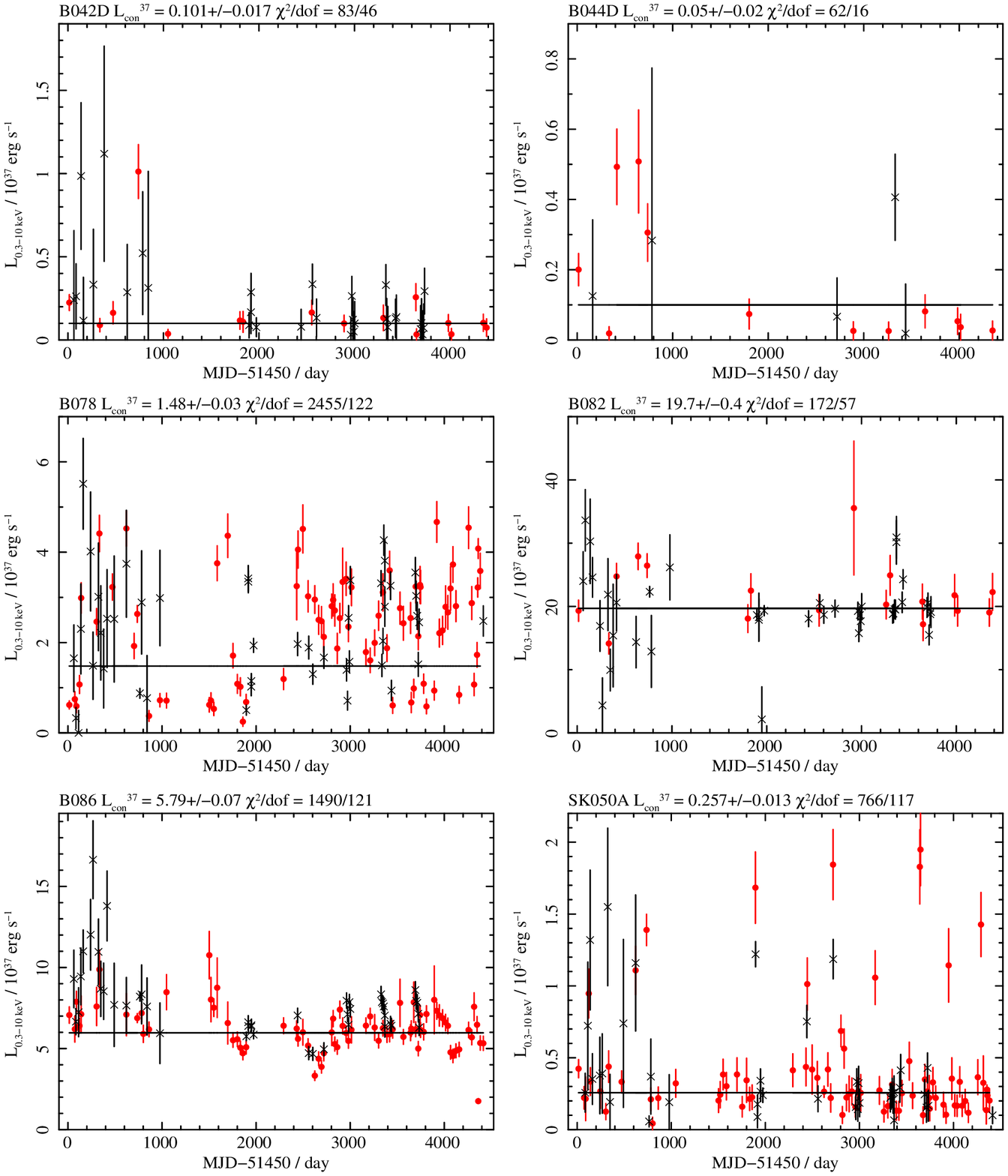}
\caption{Long-term 0.3--10 keV luminosity lightcurves of each variable X-ray source, consisting of up to  89 ACIS and 45 HRC observations. ACIS and HRC observations  are indicated by red circles and black crosses respectively. For each lightcurve we provide the best fit line of constant intensity. At the top left of each plot we list the best fit constant intensity in units of 10$^{37}$ erg s$^{-1}$ and the corresponding $\chi^2$/dof. Lightcurves for the remaining GCs are presented in the electronic edition.}\label{gclcs}
\end{figure*}

\addtocounter{figure}{-1}
\begin{figure*}
\epsscale{2.2}
\plotone{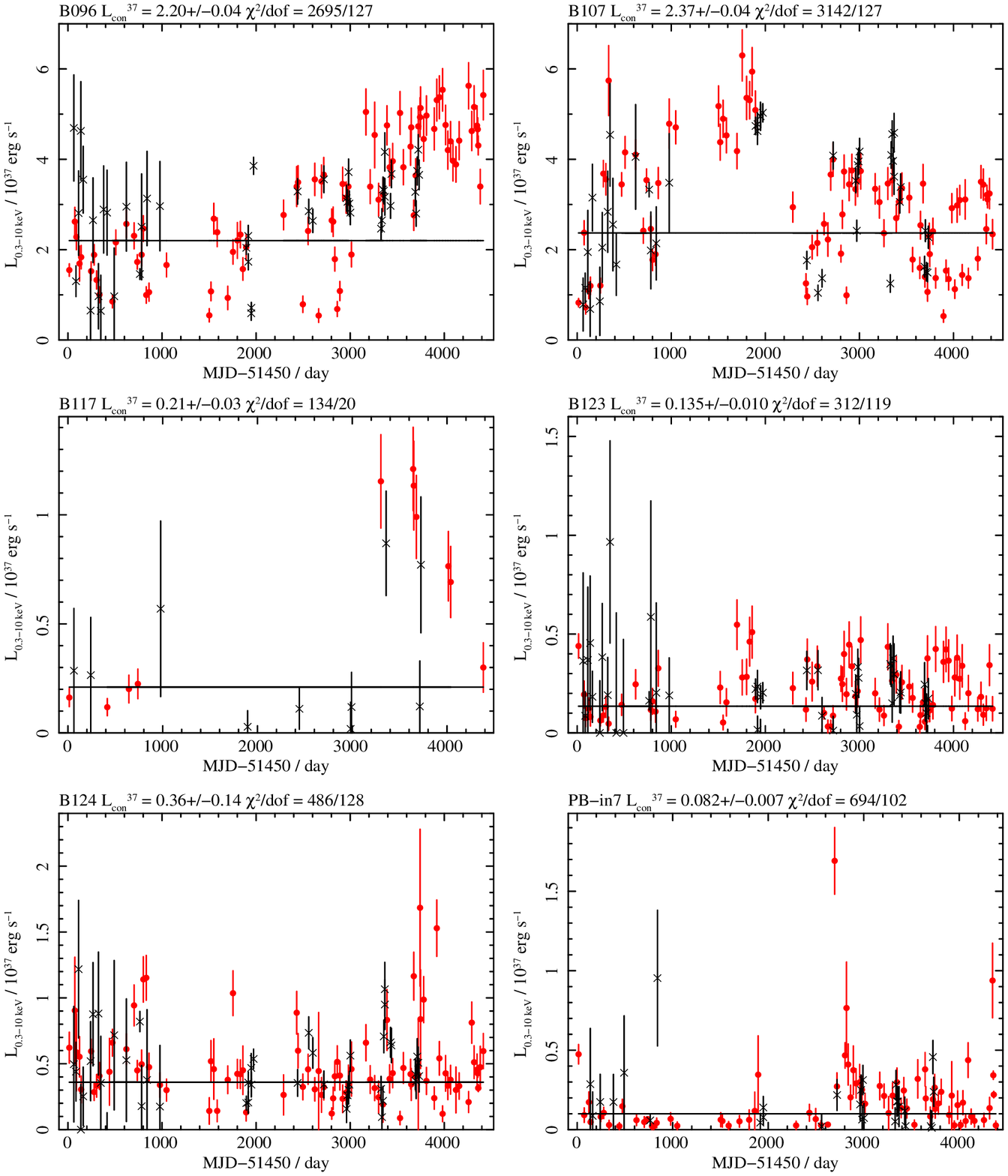}
\caption{continued}
\end{figure*}

\addtocounter{figure}{-1}
\begin{figure*}
\epsscale{2.2}
\plotone{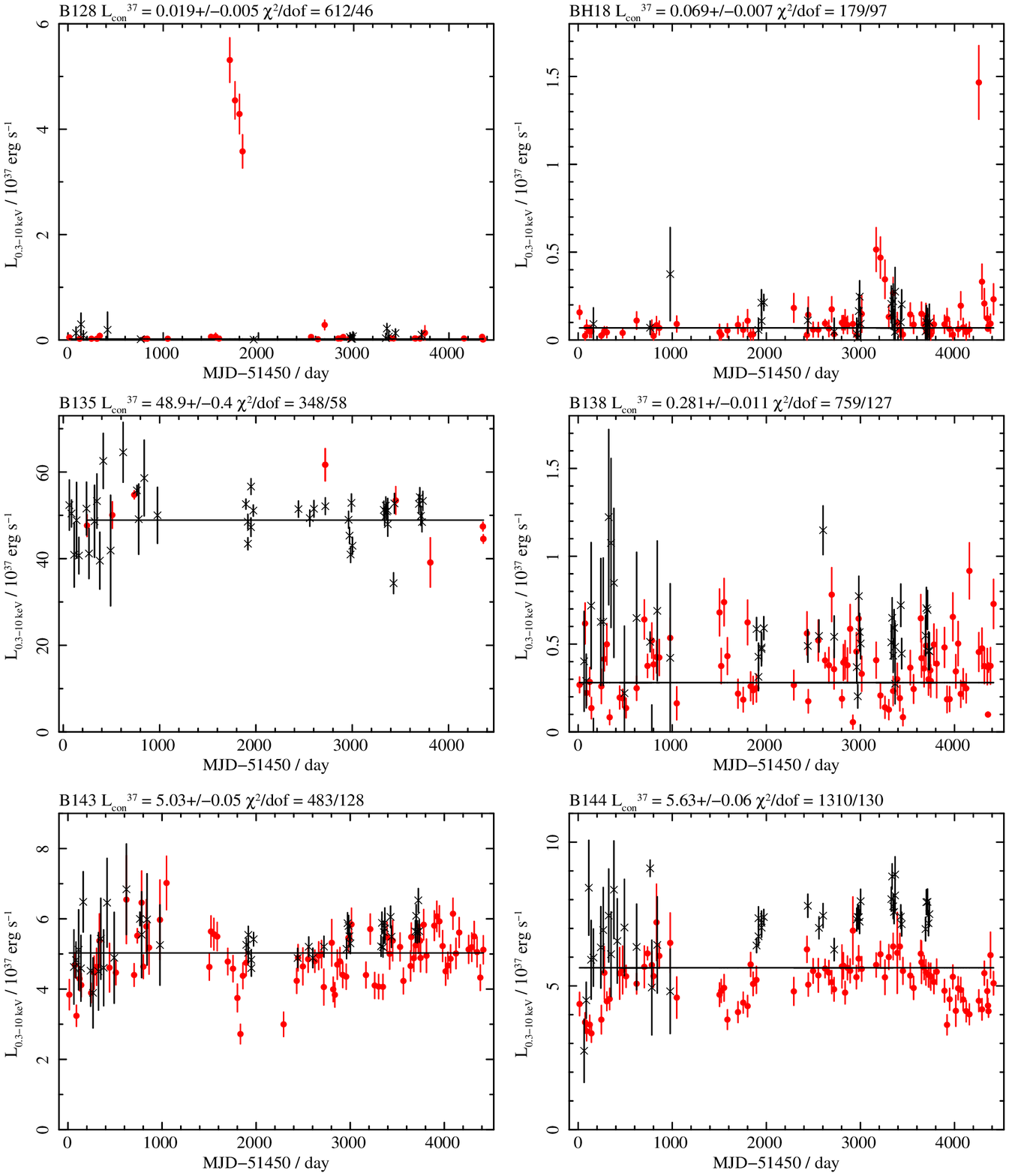}
\caption{continued}
\end{figure*}

\addtocounter{figure}{-1}
\begin{figure*}
\epsscale{2.2}
\plotone{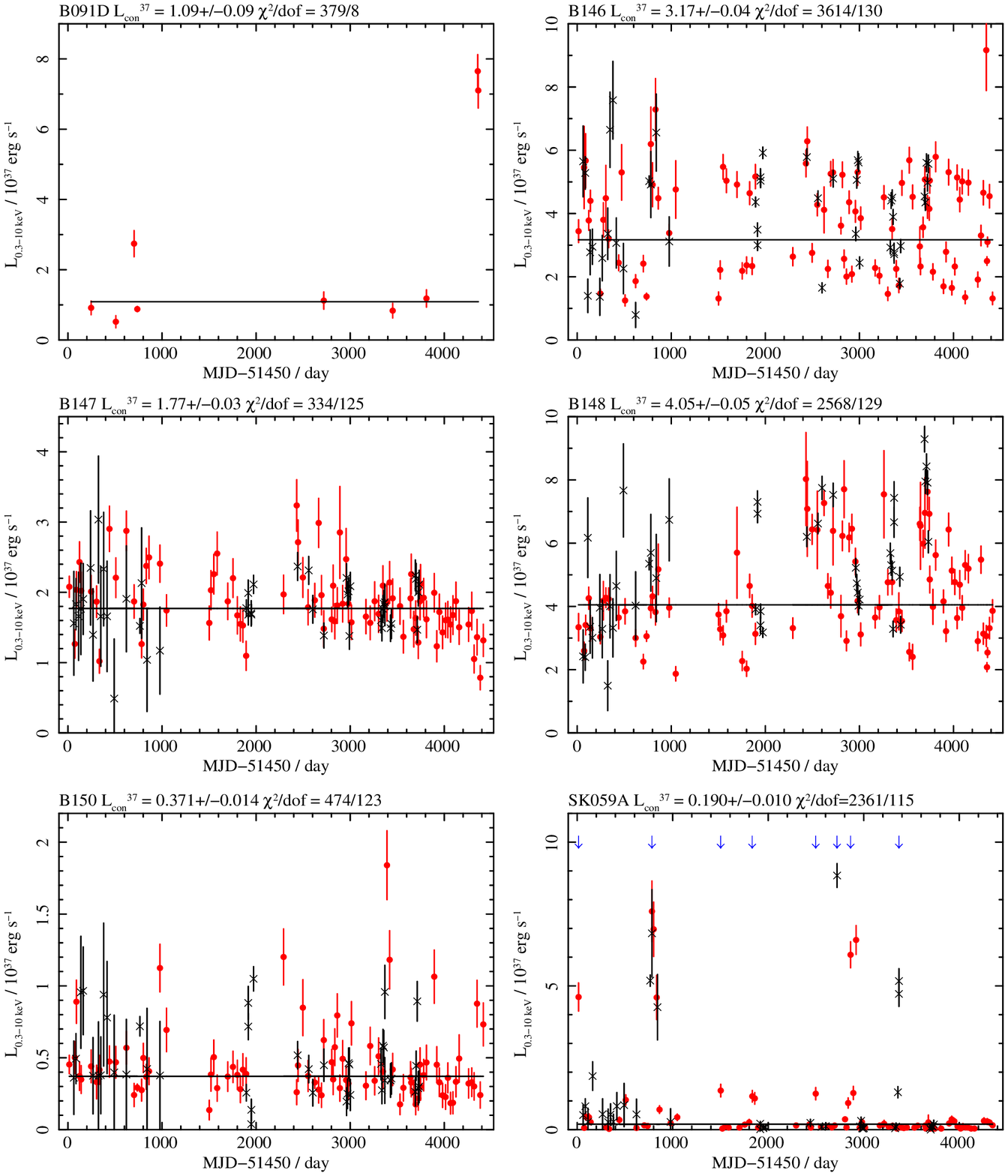}
\caption{continued. The outbursts exhibited by SK059A are indicated by downward arrows.}
\end{figure*}

\addtocounter{figure}{-1}
\begin{figure*}
\epsscale{2.2}
\plotone{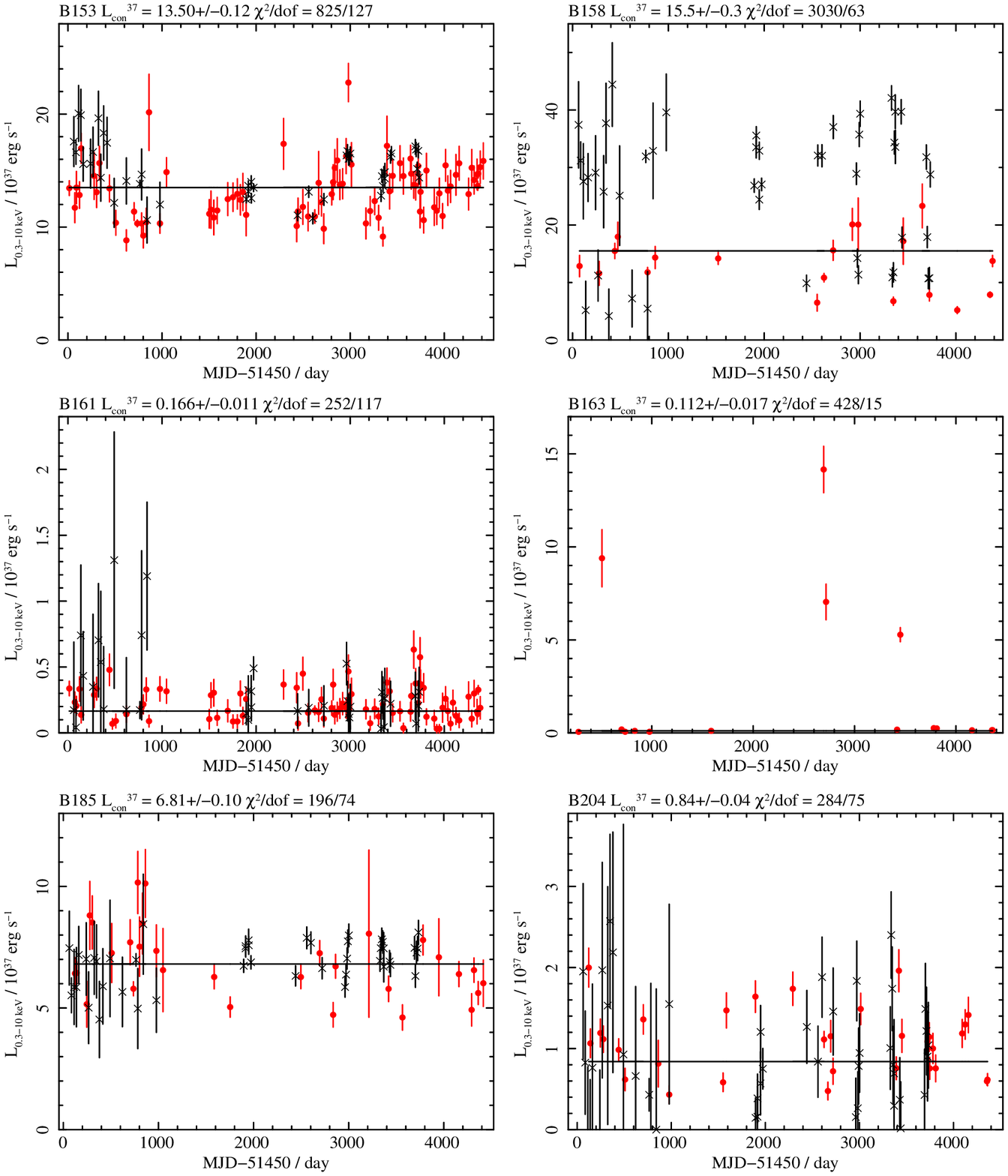}
\caption{continued }
\end{figure*}


\begin{figure*}
\epsscale{2.2}
\plotone{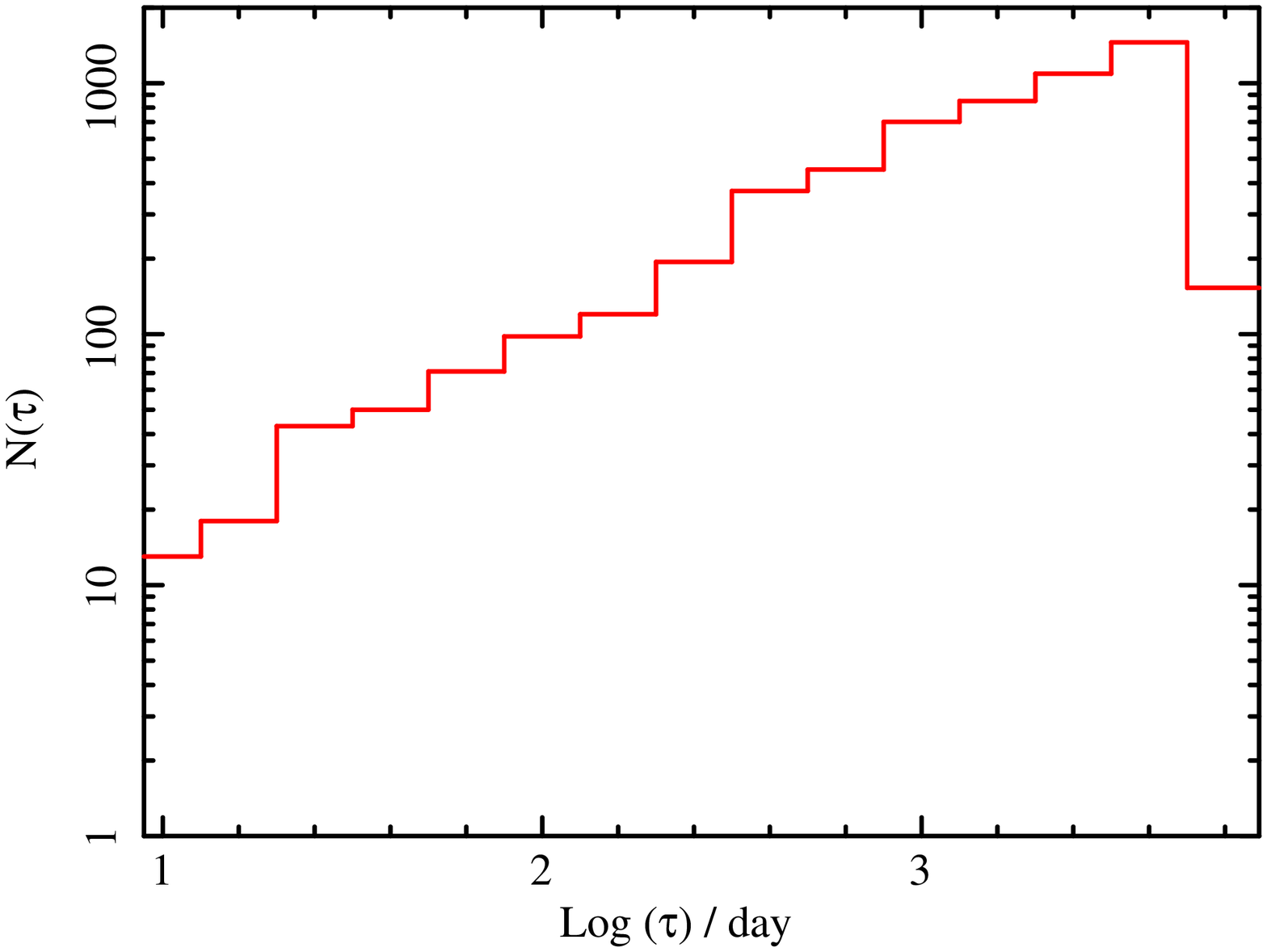}
\caption{Histogram showing the number of observations with separation $\tau$ for B144, which is the GC that was observed most often.}\label{sfh}
\end{figure*}


\begin{figure*}
\epsscale{2.2}
\plotone{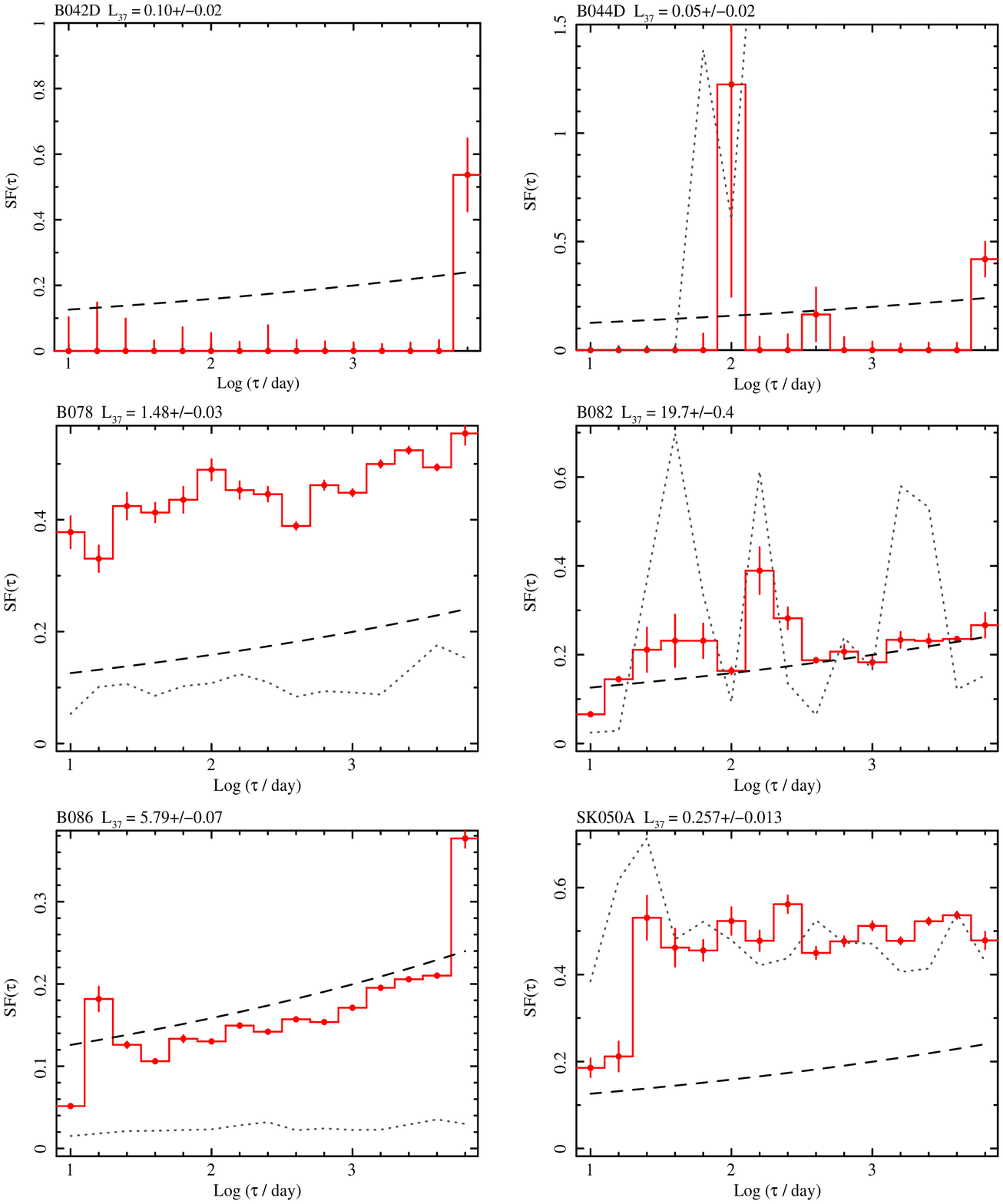}
\caption{Structure functions for each of the variable X-ray sources, created from 0.5--4.5 keV fluxes for comparison with \citet{vagnetti11}; the dashed line represents their ensemble AGN structure function. The dotted line shows $\sigma_{\rm n}^2$ for each bin for reference, since the noise component has already been subtracted from SF($\tau$). We provide the name of the X-ray source at the top of each panel, along with the best fit constant luminosity. The y axis is scaled to the SF, and high values for the noise may be excluded.  The complete figure is available online.}\label{sfs}
\end{figure*}

\addtocounter{figure}{-1}
\begin{figure*}
\epsscale{2.2}
\plotone{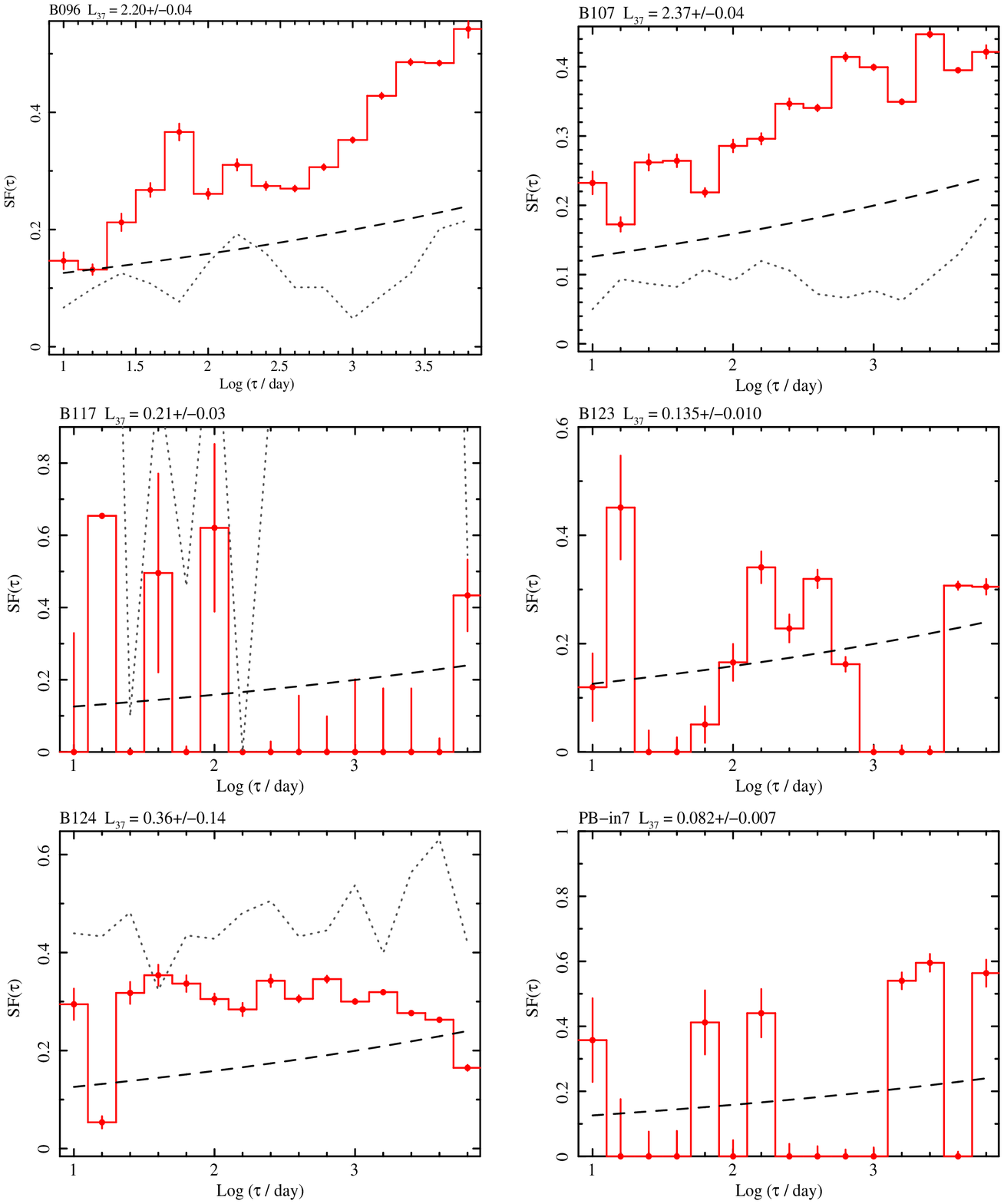}
\caption{continued}
\end{figure*}

\addtocounter{figure}{-1}
\begin{figure*}
\epsscale{2.2}
\plotone{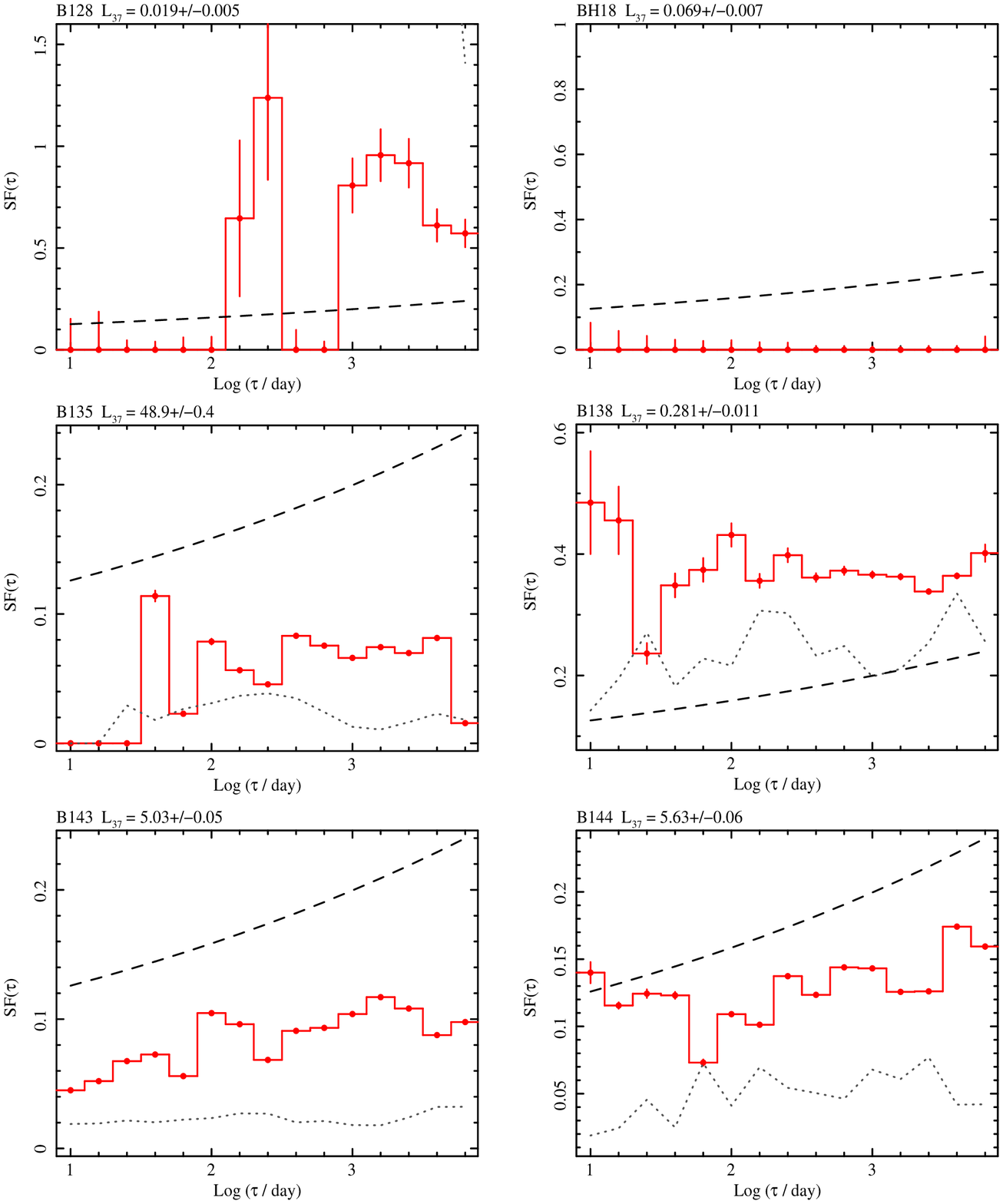}
\caption{continued}
\end{figure*}

\addtocounter{figure}{-1}
\begin{figure*}
\epsscale{2.2}
\plotone{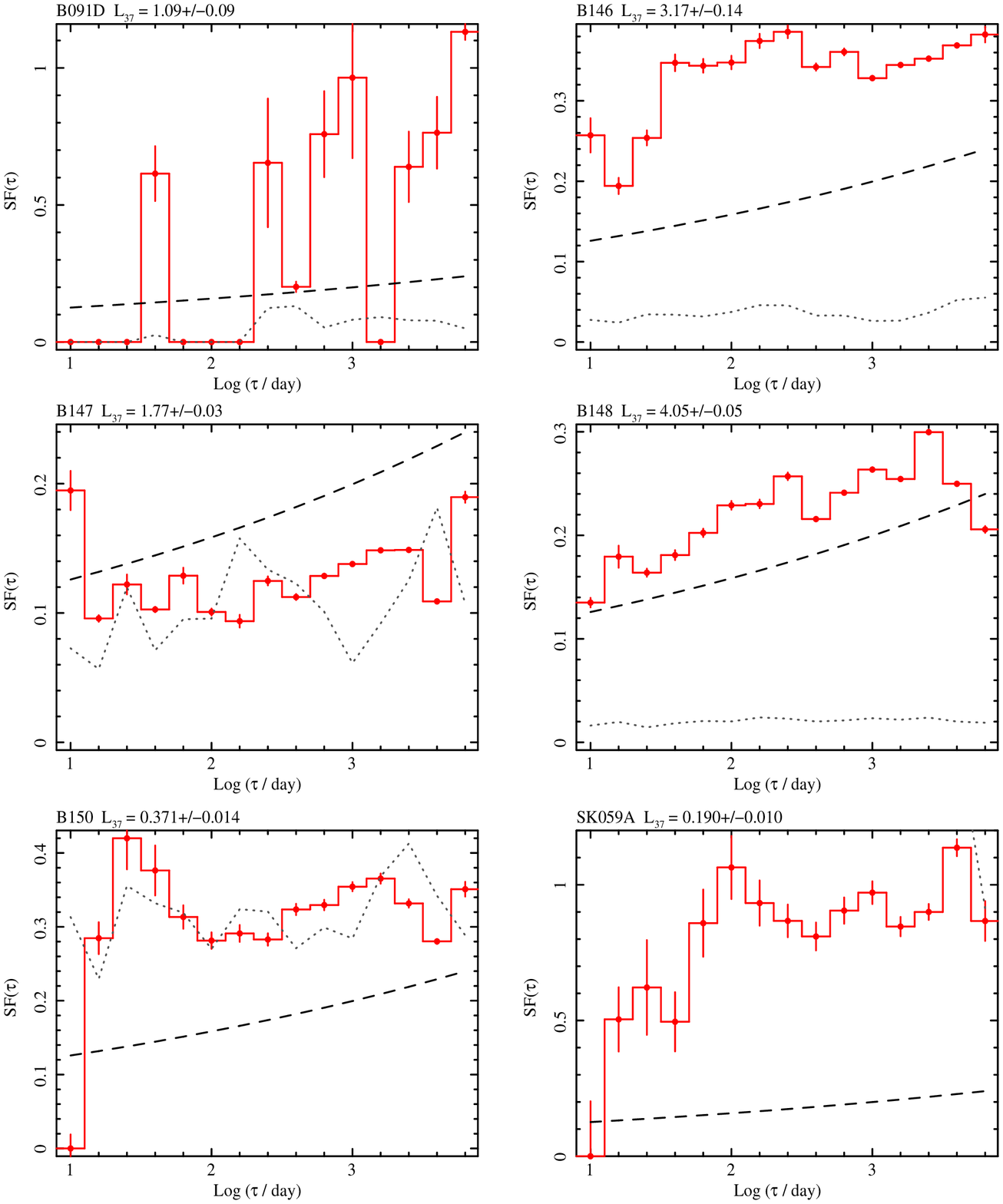}
\caption{continued}
\end{figure*}

\addtocounter{figure}{-1}
\begin{figure*}
\epsscale{2.2}
\plotone{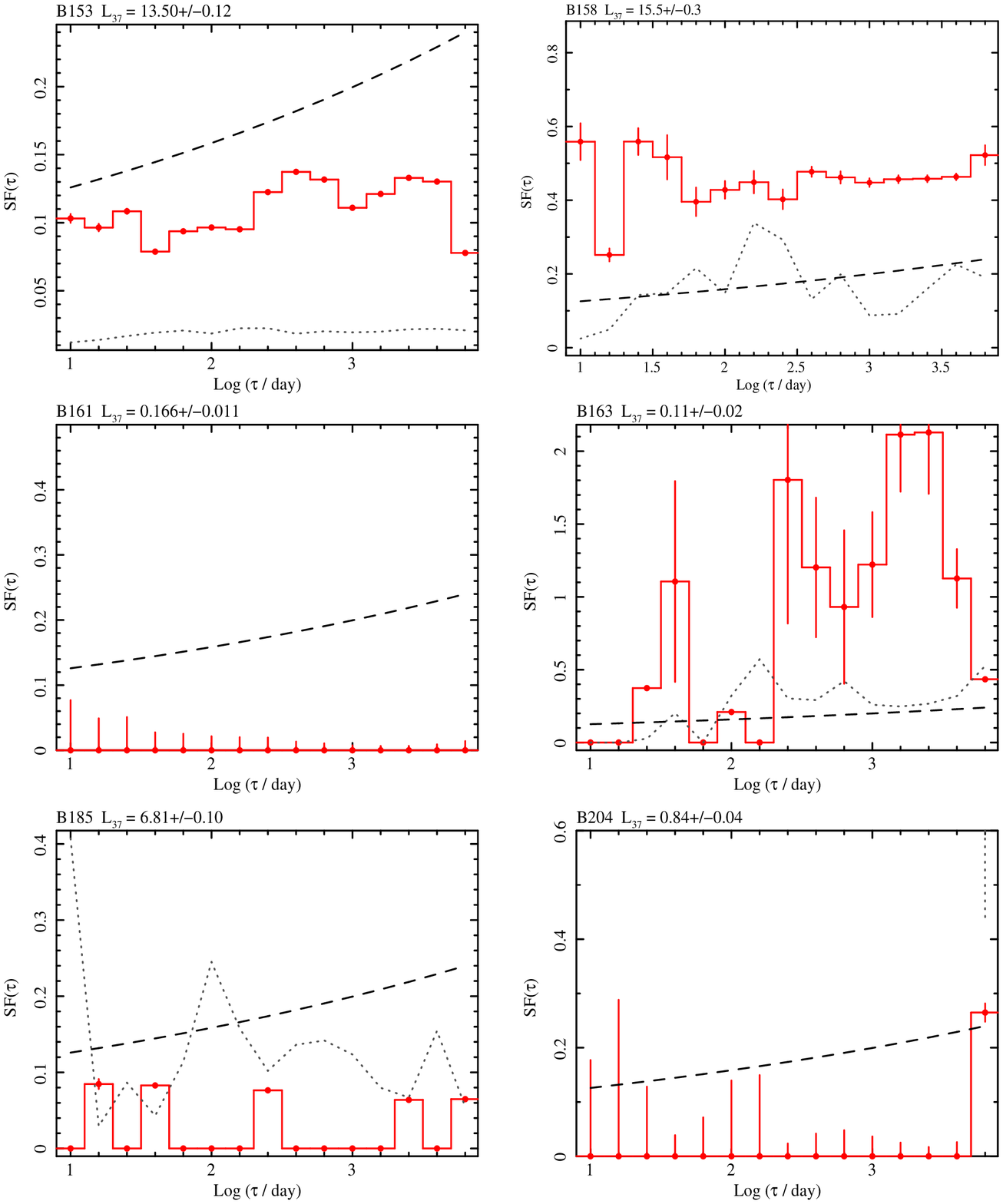}
\caption{continued }
\end{figure*}


\begin{figure*}
\epsscale{2.2}
\plotone{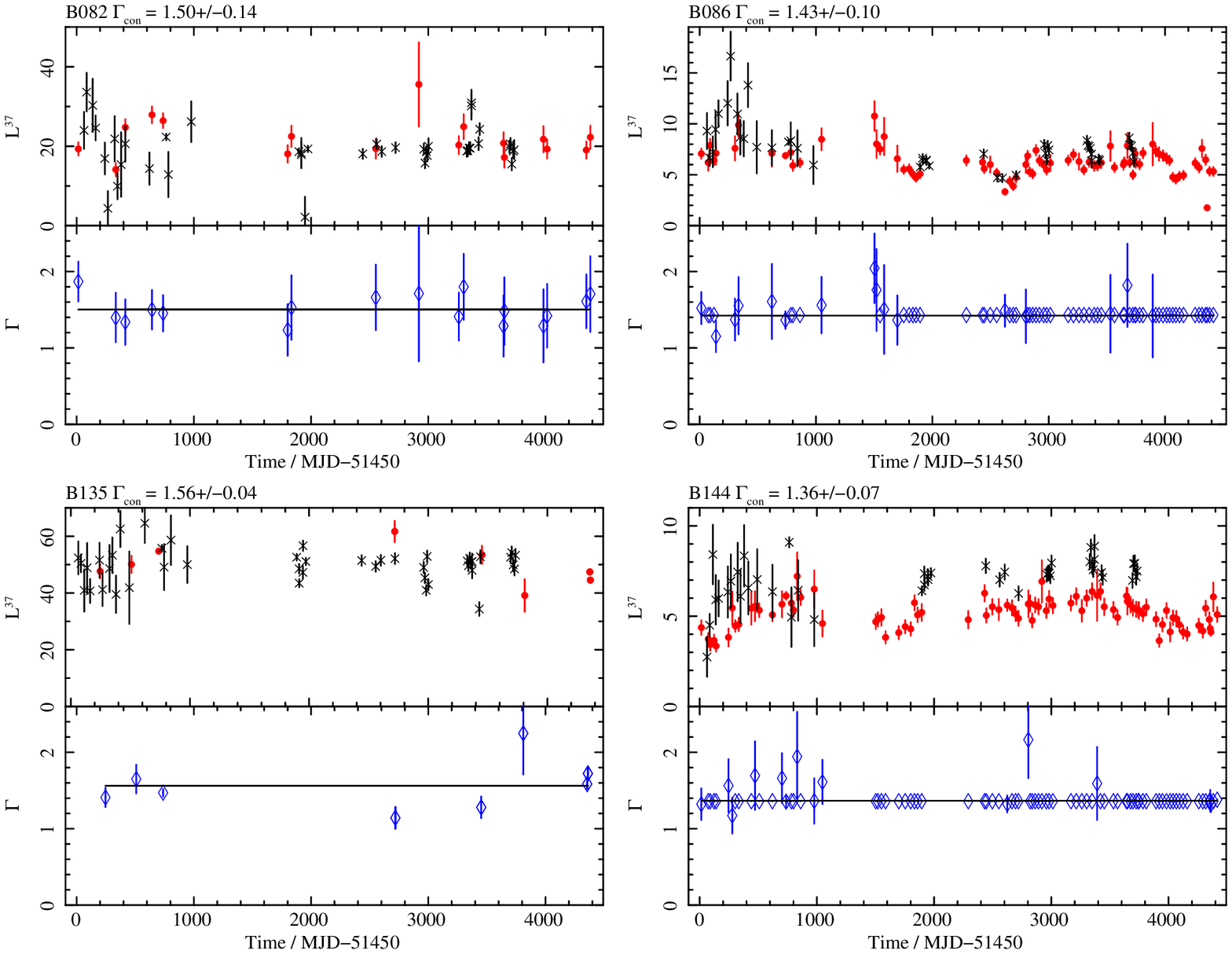}
\caption{Long term 0.3--10 keV lightcurves of black hole candidates, along with the best fit photon index, $\Gamma$,  for each ACIS observation; for observations with $<$200 net source counts $\Gamma$ is fixed to the mean $\Gamma$ for bright observations.  Luminosity uncertainties are quoted at the 1$\sigma$ level, while $\Gamma$ uncertainties are at the 90\% confidence level. The complete figure is provided in the electronic edition.}\label{bhlcs}
\end{figure*}

\addtocounter{figure}{-1}
\begin{figure*}
\epsscale{2.2}
\plotone{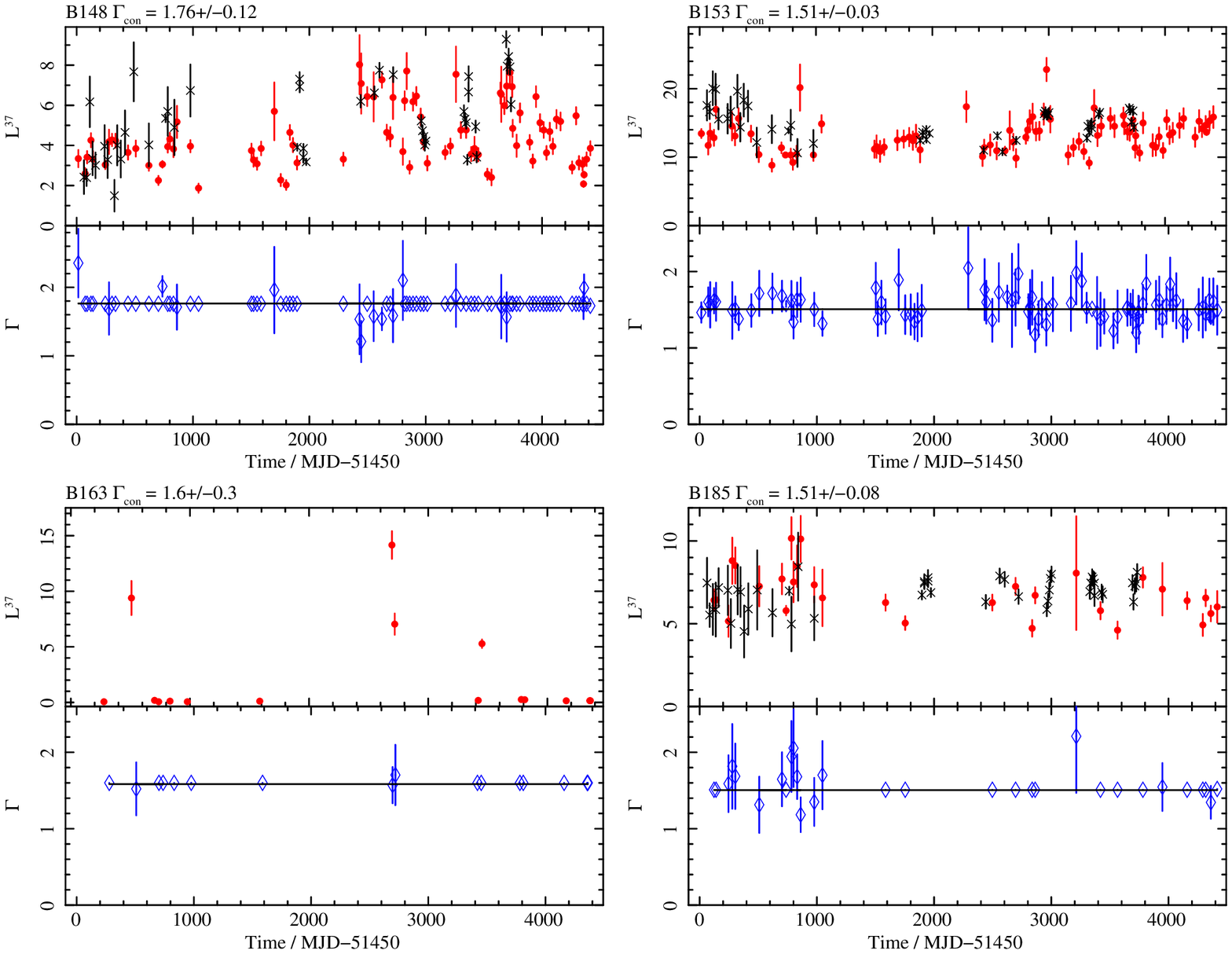}
\caption{continued}
\end{figure*}


\begin{figure*}
\epsscale{2.2}
\plotone{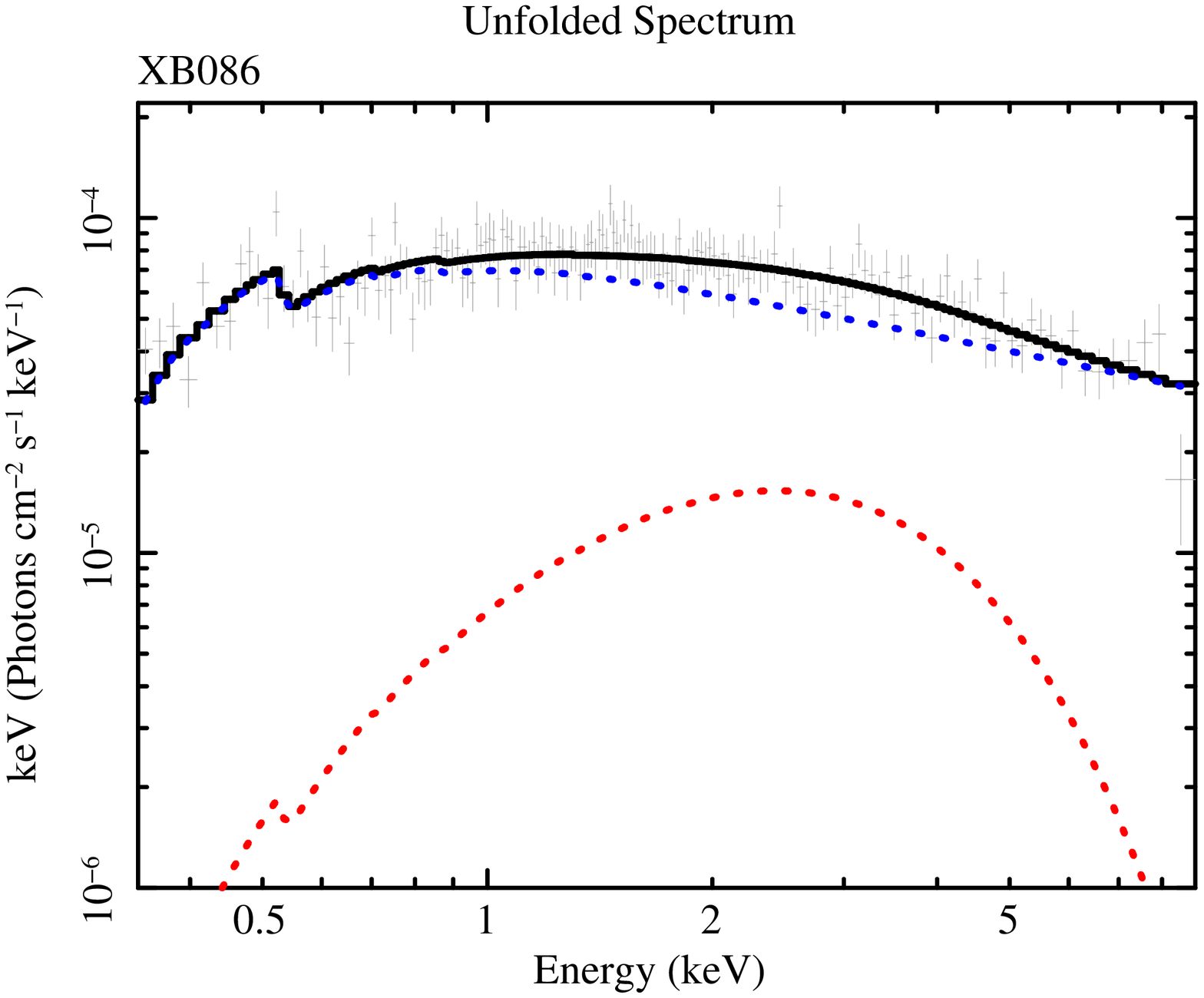}
\caption{Unfolded XMM-Newton pn spectrum for XB086, assuming the best fit absorbed blackbody + power law model; the flux of each channel is multiplied by the energy. The power law component has similar $\Gamma$ to the simple power law fit.}\label{b086spec}
\end{figure*}

\begin{figure*}
\epsscale{2.2}
\plotone{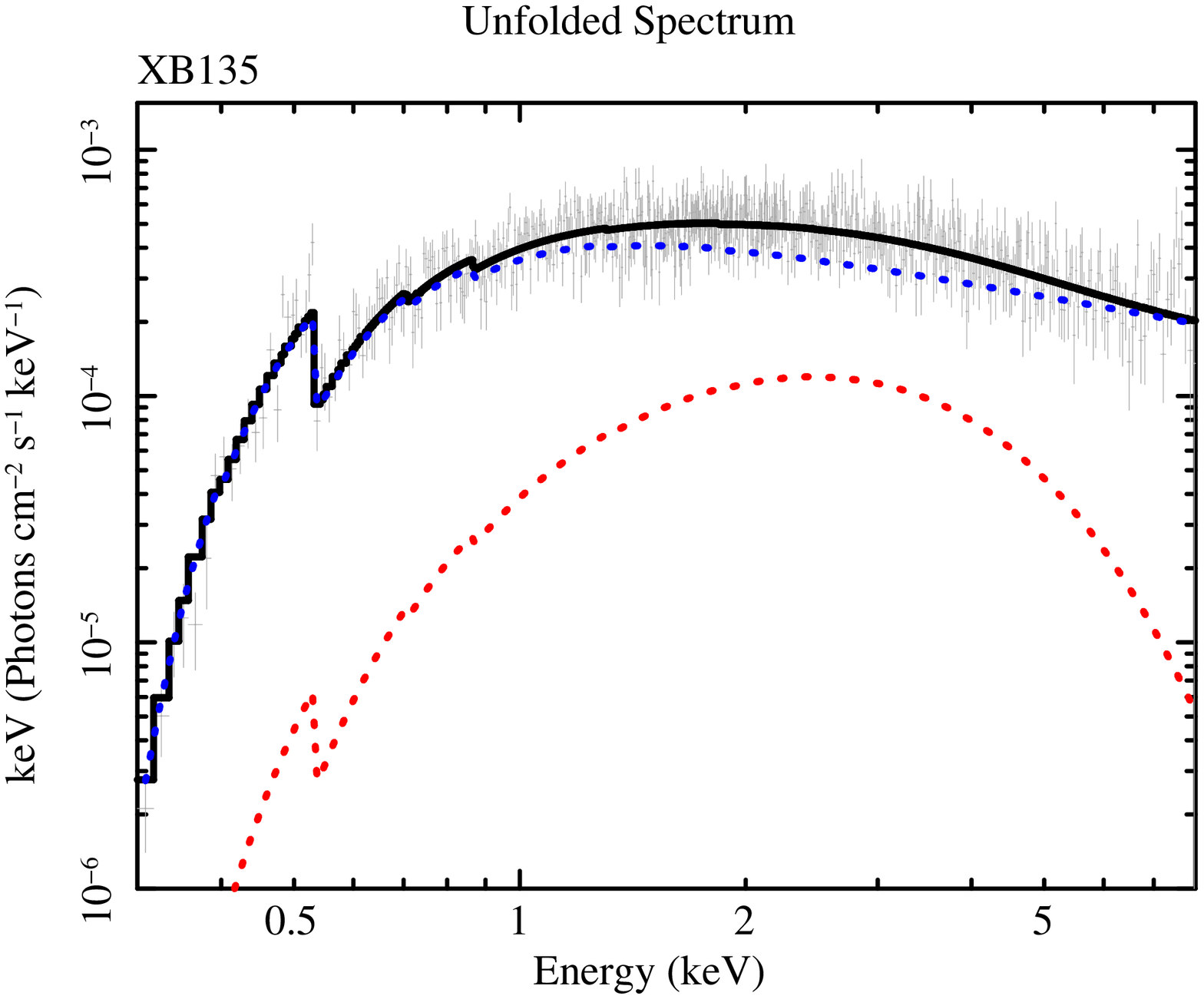}
\caption{Unfolded XMM-Newton pn spectrum for XB135, assuming the best fit absorbed blackbody + power law model; the flux of each channel is multiplied by the energy. The power law component  has similar $\Gamma$ to the simple power law fit.}\label{b135spec}
\end{figure*}

\begin{figure*}
\epsscale{2.2}
\plotone{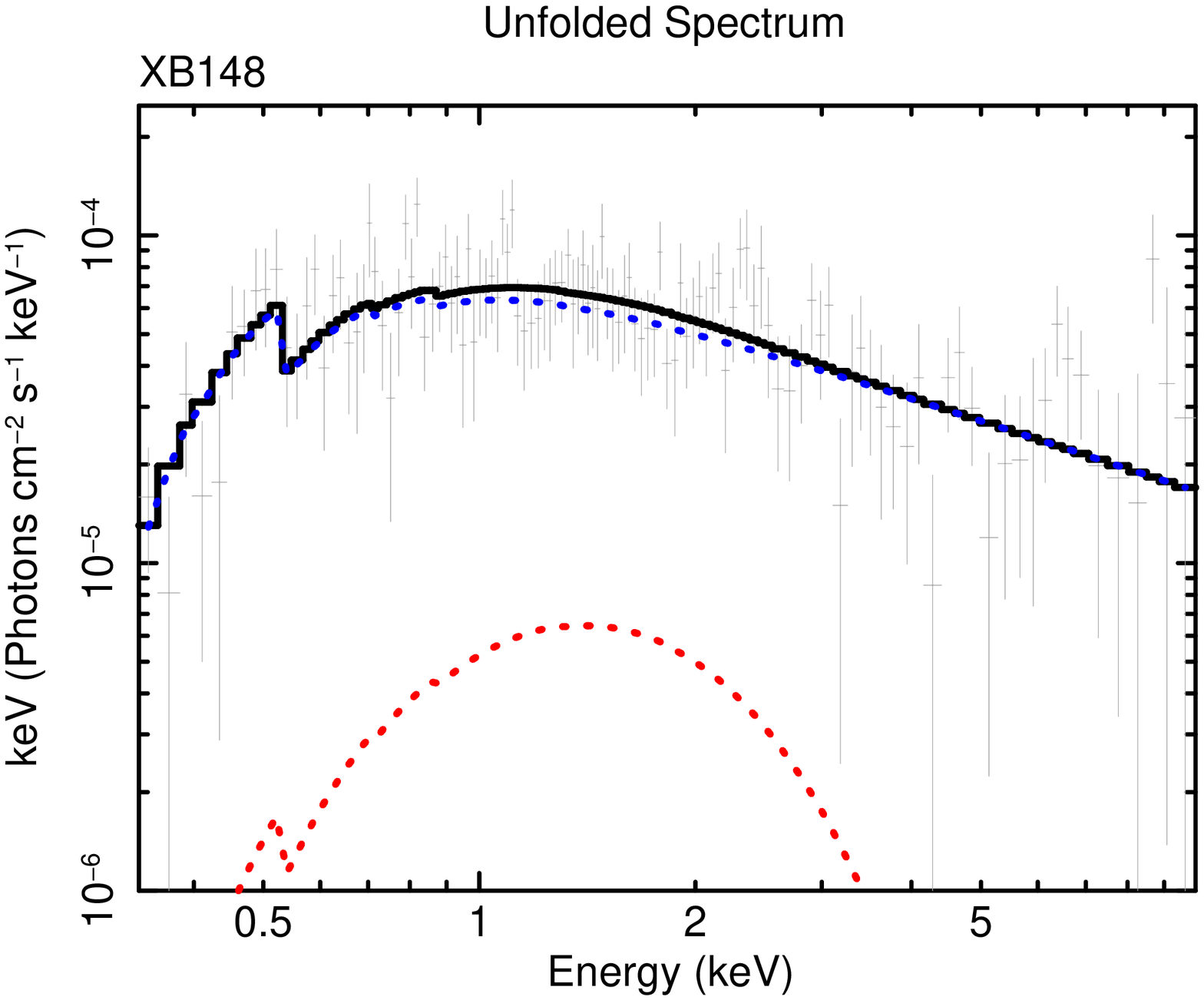}
\caption{Unfolded XMM-Newton pn spectrum for XB148, assuming the best fit absorbed blackbody + power law model; the flux of each channel is multiplied by the energy. The power law component  has similar $\Gamma$ to the simple power law fit.}\label{b148spec}
\end{figure*}

\clearpage

\begin{table*}
\begin{center}
\caption{Optical identifications for X-ray sources associated with members of the RBC V4. For each X-ray source, we provide the X-ray position, along with the name of the object from the RBC.  We then give the \citet{caldwell11} classification of the object, the distance between the X-ray source and its counterpart ($D_{\rm Obj}$), and the uncertainties in the X-ray position ( $\sigma_{\rm RA}$ and $\sigma_{\rm Dec}$). Finally we give the deviation of the X-ray position from the center of the optical counterpart (Dist/$\sigma$).  X-ray sources with no reliable  centroid solution are indicated by ``$\dots$''. } \label{opticalprops}
\begin{tabular}{cccccccccccc}
\tableline\tableline
Src &X-ray  Position & Obj & Class & $D_{\rm Obj}$/$''$ & $\sigma_{\rm RA}$/$''$ & $\sigma_{\rm Dec}$/$''$& Dist/$\sigma$\\ 
\tableline 
1 & 00:41:37.598  +40:59:20.33 & SK081B & Star & 2.6 & $\dots$ & $\dots$ & $\dots$ \\
2 & 00:42:6.011  +41:2:48.12 & B042D & Gal & 0.9 & 0.63 & 0.55 & 1.08 \\
3 & 00:42:7.070  +41:0:17.55 & B044D & Gal & 1.15 & 0.64 & 0.67 & 1.24 \\
4 & 00:42:12.198  +41:17:58.78 & B078 & Old GC & 0.54 & 0.28 & 0.27 & 1.41\\
5 & 00:42:15.847  +41:1:14.98 & B082 & Old GC & 0.68 & 0.41 & 0.37 & 1.23 \\
6 & 00:42:18.670  +41:14:02.06 & B086 & Old GC & 0.2 & 0.27 & 0.27 & 0.52 \\
7 & 00:42:19.769  +41:21:53.99 & r3-74 & Pos GC & 0.55 & 0.33 & 0.37 & 1.12 \\
8 & 00:42:21.588  +41:14:19.92 & SK050A & Pos GC & 0.38 & 0.28 & 0.27 & 0.99 \\
9 & 00:42:25.034  +40:57:19.03 & B094 & Old GC & 1.43 &3.0$^{a}$  &2.8$^{a}$  & 0.4 \\
10 & 00:42:26.074  +41:19:15.02 & B096 & Old GC & 0.23 & 0.27& 0.27 & 0.60 \\
11 & 00:42:27.392  +40:59:36.56 & B098 & Old GC & 0.69 & 3.1$^a$ & 2.5$^a$  & 0.2 \\
12 & 00:42:31.271  +41:19:38.95 & B107 & Old GC & 0.24 & 0.27 & 0.27 & 0.63 \\
13 & 00:42:33.161  +41:03:28.09& B110 & Old GC & 0.76 & 1.6$^a$ & 1.2$^a$  & 0.4 \\
14 & 00:42:34.311  +40:57:08.86 & B117 & Old GC & 0.66 & 2.2$^{a}$ &1.4$^a$ & 0.3\\
15 & 00:42:40.664  +41:10:33.67 & B123 & Old GC & 0.38 & 0.28 & 0.28 & 0.96 \\
16 & 00:42:41.451  +41:15:23.93 & B124 & Old GC & 0.26 & 0.27 & 0.27 & 0.68 \\
17 & 00:42:46.101  +41:17:36.35 & PB-in7 & Pos GC & 0.37 & 0.39 & 0.36 & 0.70 \\
18 & 00:42:47.814  +41:11:13.94 & B128 & Old GC & 0.22 & 0.27 & 0.27 & 0.58 \\
19 & 00:42:50.777  +41:10:33.80 & BH18 & Old GC & 0.71 & 0.27 & 0.27 & 1.86 \\
20 & 00:42:52.131  +41:31:05.40 & B135 & Old GC & 0.56 & 0.42 & 0.42 & 0.94 \\
21 & 00:42:55.623  +41:18:35.10 & B138 & Old GC & 0.33 & 0.61 & 0.53 & 0.41 \\
22 & 00:42:59.665  +41:19:19.25 & B143 & Old GC & 0.16 & 0.28 & 0.27 & 0.41 \\
23 & 00:42:59.872  +41:16:05.76 & B144 & Old GC & 0.23 & 0.27 & 0.27 & 0.60 \\
24 & 00:43:1.346  +41:30:17.04 & B091D & Old GC & 1.08 & 0.59$^{a}$ &0.61$^a$ & 1.3 \\
25 & 00:43:2.938  +41:15:22.60 & B146 & Old GC & 0.21 & 0.27 & 0.27 & 0.55 \\
26 & 00:43:3.309  +41:21:21.71 & B147 & Old GC & 0.21 & 0.27 & 0.27 & 0.55 \\
27 & 00:43:3.868  +41:18:04.87 & B148 & Old GC & 0.16 & 0.35 & 0.39 & 0.31 \\
28 & 00:43:7.526  +41:20:19.48 & B150 & Old GC & 0.07 & 0.44 & 0.52 & 0.10 \\
29 & 00:43:9.853  +41:19:00.67 & SK059A & HII & 0.08 & 0.43 & 0.46 & 0.13 \\
30 & 00:43:10.622  +41:14:51.37 & B153 & Old GC & 0.06 & 0.60 & 0.64 & 0.07 \\
31 & 00:43:14.388  +41:7:21.42 & B158 & Old GC & 0.28 & 1.17 & 1.35 & 0.16 \\
32 & 00:43:14.535  +41:25:12.88 & B159 & Old GC & 1.4 &2.0$^{a}$ &1.7$^a$ & 0.5 \\
33 & 00:43:15.465  +41:11:25.51 & B161 & Old GC & 0.44 & 0.59 & 0.53 & 0.56 \\
34 & 00:43:17.710  +41:27:45.53 & B163 & Old GC & 1.0 & 0.31 & 0.31 & 2.26 \\
35 & 00:43:36.798  +41:08:13.46 & B182 & Old GC & 1.9 & 2.9$^{a}$ &3.7$^a$  & 0.11\\
36 & 00:43:37.309  +41:14:43.11 & B185 & Old GC & 0.53 & 0.76 & 0.96 &0.43 \\
37 & 00:43:56.315 +41:22:01.16 & B204 & Old GC & 2.09 & 1.1$^{a}$ &1.3$^a$  & 1.2 \\

\end{tabular}
\end{center}
$^{a}$ Positional data obtained from a heavily binned image where each pixel represents 9$\times$9 native pixels. 
\end{table*}

\clearpage

\begin{table*}
\begin{center}
\caption{For each object, we provide the number of Chandra observations that include the object (ACIS and HRC). We then give the absorption ($N_{\rm H}$) and photon index  ($\Gamma$) for the model used to convert from source intensity to luminosity, followed by the $\chi^2$/dof for $N_{\rm H}$ and $\Gamma$. Finally we show the unabsorbed 0.3--10 keV luminosity and $\chi^2$/dof for the best fit line of constant intensity. Uncertainties are quoted at the 90\% confidence level.} \label{xrayprops}
\begin{tabular}{cccccccccccccc}
\tableline\tableline
Src  &  $N_{\rm ACIS}$  & $N_{\rm HRC}$ &  $N_{\rm H}$  / 10$^{20}$    &  $\Gamma$  & $\chi^2$/dof $\left( N_{\rm H} \right)$ & $\chi^2$/dof $\left( \Gamma \right)$ &   $L_{\rm con}^{37}$    &  $\chi^2$/dof $\left( L_{\rm con}^{37} \right)$ \\
\tableline 
1 & 2 & 1 & 7    & 1.7    &    &    & 0.06 $\pm$ 0.06 & 1.9/2 \\
2 & 16 & 31 & 7    & 1.7    &    &    & 0.10 $\pm$ 0.02 & 83/46 \\
3 & 12 & 5 & 7    & 1.7    &    &    & 0.05 $\pm$ 0.02 & 62/16 \\
4 & 78 & 45 & 44 $\pm$ 12 & 1.83 $\pm$ 0.18 & 0.44 / 4 & 1.5 / 4 & 1.48 $\pm$ 0.03 & 2455/122 \\
5 & 17 & 41 & 50 $\pm$ 10 & 1.5 $\pm$ 0.14 & 5 / 16 & 5 / 16 & 19.7 $\pm$ 0.4 & 172/57 \\
6 & 76 & 45 & 17 $\pm$ 7 & 1.43 $\pm$ 0.1 & 19 / 15 & 6 / 15 & 5.79 $\pm$ 0.07 & 1490/121 \\
7 & 70 & 19 & 7    & 1.7    &    &    & 0.055 $\pm$ 0.007 & 101/88 \\
8 & 79 & 39 & 21 $\pm$ 11 & 2.2 $\pm$ 0.3 &    &    & 0.257 $\pm$ 0.013 & 766/117 \\
9 & 5 & 6 & 7    & 1.7    &    &    & 0.12 $\pm$ 0.03 & 24/10 \\
10 & 83 & 45 & 42 $\pm$ 12 & 1.9 $\pm$ 0.19 & 0.6 / 3 & 4 / 3 & 2.2 $\pm$ 0.04 & 2695/127 \\
11 & 12 & 19 & 7    & 1.7    &    &    & 0.05 $\pm$ 0.02 & 33/30 \\
12 & 84 & 44 & 27 $\pm$ 9 & 1.81 $\pm$ 0.18 & 3 / 3 & 0.09 / 3 & 2.37 $\pm$ 0.04 & 3142/127 \\
13 & 19 & 29 & 7    & 1.7    &    &    & 0.15 $\pm$ 0.02 & 74/47 \\
14 & 11 & 10 & 7    & 1.7    &    &    & 0.21 $\pm$ 0.03 & 134/20 \\
15 & 77 & 43 & 7    & 1.7    &    &    & 0.135 $\pm$ 0.010 & 312/119 \\
16 & 85 & 44 & 7    & 1.7    &    &    & 0.36 $\pm$ 0.14 & 486/128 \\
17 & 75 & 28 & 7    & 1.7    &    &    & 0.082 $\pm$ 0.007 & 694/102 \\
18 & 32 & 15 & 7    & 1.7    &    &    & 0.019 $\pm$ 0.005 & 612/46 \\
19 & 73 & 26 & 7    & 1.7    &    &    & 0.069 $\pm$ 0.007 & 179/97 \\
20 & 14 & 45 & 33 $\pm$ 2 & 1.56 $\pm$ 0.04 & 66 / 13 & 50 / 13 & 48.9 $\pm$ 0.4 & 348/58 \\
21 & 85 & 43 & 7    & 1.7    &    &    & 0.281 $\pm$ 0.011 & 759/127 \\
22 & 83 & 45 & 16 $\pm$ 4 & 1.85 $\pm$ 0.12 & 1.4 / 11 & 3 / 11 & 5.03 $\pm$ 0.05 & 483/127 \\
23 & 86 & 45 & 9 $\pm$ 4 & 1.36 $\pm$ 0.07 & 6 / 13 & 7 / 13 & 5.63 $\pm$ 0.06 & 1310/130 \\
24 & 9 & 0 & 16 $\pm$ 4 & 0.74 $\pm$ 0.12 & 1.8 / 1 & 1.1 / 1 & 1.09 $\pm$ 0.09 & 379/8 \\
25 & 86 & 45 & 16 $\pm$ 5 & 2.04 $\pm$ 0.13 & 8 / 13 & 4 / 13 & 3.17 $\pm$ 0.04 & 3614/130 \\
26 & 81 & 45 & 30 $\pm$ 30 & 2 $\pm$ 0.4 &    &    & 1.77 $\pm$ 0.03 & 334/125 \\
27 & 85 & 45 & 17 $\pm$ 4 & 1.76 $\pm$ 0.12 & 3 / 15 & 12 / 15 & 4.05 $\pm$ 0.05 & 2568/129 \\
28 & 82 & 42 & 7    & 1.7    &    &    & 0.371 $\pm$ 0.014 & 474/123 \\
29 & 77 & 39 & 27 $\pm$ 16 & 1.9 $\pm$ 0.3 & 0.9 / 3 & 3 / 3 & 0.190 $\pm$ 0.010 & 2361/115 \\
30 & 83 & 45 & 12 $\pm$ 1.7 & 1.51 $\pm$ 0.03 & 24 / 79 & 35 / 79 & 13.5 $\pm$ 0.12 & 825/127 \\
31 & 19 & 45 & 26 $\pm$ 15 & 0.59 $\pm$ 0.17 & 2.4 / 8 & 4 / 8 & 15.5 $\pm$ 0.3 & 3030/63 \\
32 & 23 & 24 & 7    & 1.7    &    &    & 0.069 $\pm$ 0.011 & 48/46 \\
33 & 81 & 37 & 7    & 1.7    &    &    & 0.166 $\pm$ 0.011 & 252/117 \\
34 & 15 & 0 & 12 $\pm$ 12 & 1.6 $\pm$ 0.3 & 0.11 / 2 & 0.12 / 2 & 0.11 $\pm$ 0.02 & 428/15 \\
35 & 9 & 20 & 7    & 1.7    &    &    & 0.049 $\pm$ 0.010 & 26/28 \\
36 & 30 & 45 & 11 $\pm$ 4 & 1.51 $\pm$ 0.08 & 10 / 25 & 13 / 25 & 6.81 $\pm$ 0.10 & 196/74 \\
37 & 31 & 45 & 7    & 1.7    &    &    & 0.84 $\pm$ 0.04 & 284/75 \\

\tableline 
\end{tabular}
\end{center}
\end{table*}

\clearpage

\begin{table*}
\begin{center}
\caption{Ages, masses and metallicities  (expressed as [Fe/H]) of the GCs hosting our BHCs, taken from \citet{caldwell11}. We also rank the GCs amongst the 379 GCs studied by \citet{caldwell11} by mass and metallicity.  } \label{gcprops}
\begin{tabular}{lcrrrr}
\tableline\tableline
GC & Log Age/ yr & Log Mass/$M_{\odot}$  &  Mass Rank & [Fe/H] &  [Fe/H] Rank \\ 
\tableline 
B045 & 10.2 & 5.94 &  69 ($>$82\%) & $-$0.86$\pm$0.11 &  137 ($>$64\%) \\
B082 & 10.1 & 6.72 &  4 ($>$99\%) & $-$0.74$\pm$0.11 &  102 ($>$73\%) \\
B086 & 10.2 & 6.13 &  33 ($>$91\%) & $-$1.82$\pm$0.16 &  294 ($>$22\%) \\
B135 & 10.2 & 5.97 &  64 ($>$83\%) & $-$1.82$\pm$0.13 &  296 ($>$22\%) \\
B144 & 10.1 & 5.58 &  139 ($>$63\%) & +0.08$\pm$0.17 &  10 ($>$97\%) \\
B148 & 10.2 & 6.04 &  50 ($>$87\%) & $-$1.02$\pm$0.11 &  181 ($>$52\%) \\
B153 & 10.1 & 5.81 &  87 ($>$77\%) & $-$0.28$\pm$0.13 &  36 ($>$91\%) \\
B163 & 10.1 & 6.26 &  20 ($>$95\%) & $-$0.13$\pm$0.11 &  21($>$94\%) \\
B185 & 10.1 & 6.03 &  53 ($>$86\%) & $-$0.61$\pm$0.11 &  75 ($>$80\%) \\
\end{tabular}
\end{center}
\end{table*}

\end{document}